%% file: CPdual_final_mod_2.tex
\documentclass[a4paper,12pt,notitlepage]{article}
\usepackage[dvips]{graphicx}
\usepackage{amsfonts}
\usepackage{amsmath,amsthm}
\usepackage{mathrsfs}
\usepackage{makeidx}
\usepackage{xypic}
\usepackage{pstricks}
\usepackage{pst-node}
\usepackage{url}
\usepackage{multirow}

\usepackage[arrow, matrix, curve]{xy}
\xymatrixcolsep{0.5cm}
\xymatrixrowsep{0.5cm}
\numberwithin{equation}{section}

\DeclareMathOperator{\Cas}{Cas} \DeclareMathOperator{\gl}{u}
\DeclareMathOperator{\psl}{psl} \DeclareMathOperator{\OSp}{osp}
\DeclareMathOperator{\OO}{O} \DeclareMathOperator{\UU}{U}
\DeclareMathOperator{\End}{End} \DeclareMathOperator{\Hom}{Hom}
\DeclareMathOperator{\ssl}{sl}

\newcommand{\beqa}{\begin{eqnarray}}
\newcommand{\eeqa}{\end{eqnarray}}
\newcommand{\beq}{\begin{equation}}
\newcommand{\eeq}{\end{equation}}
\DeclareMathOperator{\str}{\text{str}}

\newcommand\Ssn{\ensuremath{S^{2S-1|2S}}}

\newcommand\Un{\ensuremath{\text{U}(S\,|S)}}
\newcommand{\U}[2]{\ensuremath{\text{U}(#1\,|#2)}}

\newcommand\CPn{\ensuremath{\text{$\mathbb{CP}$}^{S-1|S}}}
\newcommand\CPone{\ensuremath{\text{$\mathbb{CP}$}^{1|2}}}
\newcommand\CPzero{\ensuremath{\text{$\mathbb{CP}$}^{0|1}}}
\newcommand\CPthree{\ensuremath{\text{$\mathbb{CP}$}^{3|4}}}
\newcommand\bCPn{\ensuremath{\text{$\mathbb{CP}$}^{S}}}
\newcommand\bCPone{\ensuremath{\text{$\mathbb{CP}$}^{1}}}
\newcommand\tr{{\text{tr}}}

\newcommand\psutwo{\ensuremath{\text{psu}(2|2)}}
\newcommand\putwo{\ensuremath{\text{pu}(2\,|2)}}
\renewcommand\u{\ensuremath{\text{u}(S\,|S)}}
\newcommand\un{\ensuremath{\text{u}(S\,|S)}}
\newcommand\utwo{\ensuremath{\text{u}(2|2)}}
\newcommand\Utwo{\ensuremath{\text{u}(2|2)}}
\def\pl{\partial}
\newcommand\sltwo{\ensuremath{\text{sl}_2}}
\newcommand\uone{\ensuremath{\text{u}(1)}}
\newcommand\uoneone{\ensuremath{\text{u}(1|1)}}

\newcommand{\bracket}[2]{\ensuremath{\left< #1\, ,\, #2\right>}}
\begin{document}
\title{\textbf{The Sigma Model on \\
Complex Projective Superspaces}}

\author{\\[5mm]Constantin Candu$^1$, Vladimir Mitev$^1$, Thomas
Quella$^2$,\\[2mm]
Hubert Saleur$^{3,4}$ and Volker Schomerus$^1$\\[5mm]
$^1$ DESY Hamburg, Theory Group, \\
Notkestrasse 85, D--22607 Hamburg, Germany
\\[3mm]
$^2$ Institute for Theoretical Physics, University of Amsterdam,\\
Valckenierstraat 65, 1018 XE Amsterdam, The Netherlands\\[3mm]
$^3$ Institute de Physique Th\'eorique, CEA Saclay,\\  F-91191
Gif-sur-Yvette, France\\[3mm]
$^4$ Physics Dept., USC, Los Angeles, CA 90089-0484, USA}
\pagestyle{headings}
\date{}

\begin{titlepage}
    \maketitle
    \begin{abstract}
    The sigma model on projective superspaces \CPn\ gives rise
    to a continuous family of interacting 2D conformal field
    theories which are parametrized by the curvature radius $R$
    and the theta angle $\theta$. Our main goal is to determine
    the spectrum of the model, non-perturbatively as a function
    of both parameters. We succeed to do so for all open
    boundary conditions preserving the full global symmetry
    of the model. In string theory parlor, these correspond
    to volume filling branes that are equipped with a monopole
    line bundle and connection. The paper consists of two parts.
    In the first part, we approach the problem within the
    continuum formulation. Combining combinatorial arguments with
    perturbative studies and some simple free field calculations,
    we determine a closed formula for the partition function
    of the theory. This is then tested numerically in the
    second part. There we extend the proposal of
    [\url{arXiv: 0908.1081}] for a spin chain regularization of
    the \CPn\ model with open boundary conditions and use it
    to determine the spectrum at the conformal fixed point.
    The numerical results are in remarkable agreement with the
    continuum analysis.
    \end{abstract}
\vspace*{-21cm} {\tt {DESY 09-120}} \hfill {\tt {0908.0878}}
\end{titlepage}

\addtolength{\baselineskip}{3pt}
\section{Introduction}

Sigma models with target space supersymmetry are of much recent
interest. They possess a number of truly remarkable properties.
Most importantly, many of them come in continuous families with
vanishing $\beta$-function, i.e.\ they provide examples of
(non-unitary) 2-dimensional conformal field theories with
continuously varying exponents. There are several series of such
models that arise from compact symmetric superspaces, including
the sigma model on the odd-dimensional superspheres $S^{2S+1|2S}$
and on the complex projective superspaces \CPn. Their systematic
investigation was initiated in \cite{Read:2001pz}. More recently,
the superspheres have been reconsidered both through numerical and
algebraic investigations of lattice discretizations
\cite{Candu:2008vw,Candu:2008yw} and within the continuum
formulation \cite{Mitev:2008yt}. In particular, it was shown that
the conformal weights of fields with open boundary conditions
possess a very simple dependence on the curvature radius of the
supersphere. In fact, when Neumann boundary conditions are imposed
on both ends of the strip, the boundary partition function
can be computed exactly. The resulting formula provides strong evidence
for a remarkable new non-perturbative (in the radius $R$ or,
equivalently, the sigma model coupling $g_\sigma$) duality between
the sigma model on superspheres and the OSP($2S+2|2S$) Gross-Neveu
model. It generalizes the famous duality between the compactified
free boson and the massless Thirring model.
\smallskip

The aim of this note is to extend the investigations of
\cite{Candu:2008vw,Mitev:2008yt} to sigma models with target space
\CPn. The sigma model on complex projective superspaces gives rise
to a 2-parameter family of conformal fields theories with central
charge $c=-2$. In addition to the sigma model coupling $g_\sigma$
(or radius $R$) one can also introduce a theta term with arbitrary
coefficient $\theta$. There are several reasons to be interested
in these models. To begin with, the spaces \CPn\ are the simplest
examples of Calabi-Yau supermanifolds. Supersymmetric sigma models
on these spaces have been investigated by several authors (see
e.g.\ \cite{Aganagic:2004yh,Kumar:2004dj,Ahn:2004xs,Belhaj:2004ts,
Ricci:2005cp,Seki:2006cj}) after Witten had proposed the open
topological B-model on \CPthree\ as a candidate for a string
theoretic description of N=4 super Yang-Mills theory
\cite{Witten:2003nn}. Further motivation comes from one of the
ramifications of the $AdS/CFT$ correspondence. According to a
recent conjecture in \cite{Aharony:2008ug}, the IR fixed point
of the effective gauge field theory on a stack of D2 branes is
dual to string theory on $AdS_4 \times \CPthree$. Though
our findings for the sigma model on \CPn\ do not possess direct
applications to the spectrum of string theory, our study throws
light on some basic features such as e.g. the issue of instanton
corrections to physical quantities. Since the bosonic base of
\CPn\ has a non-trivial second cohomology $H_2$, world-sheet
instanton solutions to the classical sigma model do exist and
signal the possibility of non-perturbative effects. These seem
well worth a more detailed investigation, both in the \CPn\
sigma models and the string background $AdS_4 \times \CPthree$.
A third motivation we want to mention is related to the theory
of quantum Hall plateau transitions. The model we are about
to study may be considered as a compact relative of the sigma
model $\ensuremath{\text{U}(1,1|2)}/\ensuremath{\text{U}(1|1)}
\times \ensuremath{\text{U}(1|1)}$ at $\theta=\pi$
%
that has been extensively studied in this context
in \cite{Zirn_prev, Zirnbauer}. Some of our
constructions and results suggest interesting extensions to
the non-compact model. We shall come back to the last two
applications in the concluding section.

Let us now describe the content of the paper and its main results.
Our work is split into two parts, one on the continuum
formulation, the other on numerical studies of a lattice
discretization. Our presentation begins with a review of the
classical sigma model on \CPn. Since we are mostly interested in
world-sheets with boundaries, particular attention will be paid to
boundary conditions.  In particular, we analyze the possible \Un\
symmetric boundary conditions. As we shall discuss, these are
associated with complex line bundles on \CPn\ and hence they are
labelled by an integer $M$. In Section 3 we analyze the quantum
theory in the limit of infinite radius $R$ (vanishing sigma model
coupling $g_\sigma$). We start our presentation by reviewing the
state space of a particle on \CPone\ in the presence of a monopole
gauge field. One of the two different descriptions we provide
generalizes straightforwardly to the full quantum field theory.
The main goal of section 3 is to construct and analyze the
partition function \eqref{completespectrum2} of the boundary field
theory at $R = \infty$. Our strategy then is to deform the
partition function from $R = \infty$ to finite $R$. To this end we
adapt the background field expansion to the sigma model on \CPn
and explain how to compute boundary 2-point functions. As in the
case of superspheres, there are remarkable cancellations in the
expansion for boundary conformal weights. These suggest an exact
expression \eqref{ZRfinite} for the partition function of the
model at finite radius and for arbitrary value of the $\theta$
angle. After explaining the various ingredients of this formula,
we extract a list of consequences that will be confronted with
numerical tests.

In the second part, we consider a lattice version of the sigma
model on \CPn.
This lattice version was first studied in \cite{Candu}.
Our approach does not rely on direct discretization
of the action and Monte Carlo simulations, but rather on the
general relation between sigma models and spin chains. We are thus
led to study an "antiferromagnetic" spin chain where the degrees
of freedom  take values in an alternating sequence of \u\  modules
$V$ and $V^\star$.  By allowing interactions between nearest
neighbors as well as second nearest neighbors, we are able to
recover the spectrum of the \CPn\ model, and to check most
predictions from the continuum theory. In particular, we find good
agreement with our proposal for the exact partition function
\eqref{ZRfinite}, and we determine the running coupling constant
$g_\sigma^2$ in terms of the lattice parameters. We also come up
with a natural lattice version of the boundary conditions of the
sigma model associated with non trivial complex line bundles.

\newpage

\noindent
{\huge{\bf Part I: Continuum Theory}}\\[2cm]
In the first part we shall approach the \CPn\ model through its
continuum formulation. Target space supersymmetry will allow us to
find exact expressions for the conformal weights of boundary
fields as a function of radius $R$ and theta angle $\theta$.

\section{The Sigma Model on Projective Superspace}

The aim of this section is to review some facts about the complex
projective superspace \CPn\ and the non-linear sigma model
thereon. In the first subsection we discuss two different
formulations of the theory. The first one involves a constraint
and it is manifestly \Un\ invariant. There exists an alternative
description, in which the constraint is solved at the expense of
breaking the \Un\ symmetry down to U($S-1|S$). Both formulations
will play some role in the subsequent analysis. The second
subsection contains a comprehensive analysis of \Un\ symmetric
boundary conditions. We shall argue that there exists an infinite
family of such boundary conditions, one for each integer $M$. They
correspond to the choice of a complex line bundle in \CPn\ along
with a connection one-form $A_M$. For $S=2$ the connection
one-form is a supersymmetric version of the gauge field produced
by a Dirac monopole of charge $M$.

\subsection{The sigma model on \CPn}

Complex projective superspaces \CPn\ are built in a way that
resembles closely the construction of their bosonic cousins.
We begin with flat superspace $\mathbb{C}^{S|S}$. The $S$ complex
bosonic coordinates are denoted by $z_a$ and we use $\xi_a$ for
the $S$ fermionic directions. Within the flat complex superspace
we consider the odd (real) dimensional supersphere defined by the
equation
\begin{equation} \sum_{a=1}^S z_a z_a^\ast + \sum_{a=1}^S \xi_a
\xi_a^\ast = 1 \ \ . \label{SSconstraint}
\end{equation}
The supersphere \Ssn\ carries an action of U(1) by simultaneous phase 
rotations of all bosonic and fermionic coordinates,
\begin{equation}\label{phrot}
 z_a \ \longrightarrow \ e^{i \varpi} z_a \ \ \ \ \ , \ \ \ \
   \xi_a \ \longrightarrow\ e^{i\varpi} \xi_a \ \ .
\end{equation}
Note that this transformation indeed leaves the constraint invariant. 
The complex projective superspace \CPn\ is the quotient space
\Ssn\!/U(1).
\smallskip

Functions on the supersphere \Ssn\ carry an action of the the Lie
supergroup \Un. These transformations include the phase rotations
\eqref{phrot} which act trivially on \CPn. Hence, the stabilizer
subalgebra of a point on the projective superspace is given by
u(1) $\times$ u($S-1|S$) where the first factor corresponds to the
action \eqref{phrot}. We conclude that
\begin{eqnarray}
\CPn \ = \ \text{\Un} /\left(\,\text{U(1) $\times$
U($S-1|S$)}\right) \ \ .
\end{eqnarray}
Their simplest representative is \CPzero\, i.e. the space with
just two real fermionic coordinates. The sigma model with this
target space is equivalent to the theory of two symplectic
fermions, which has been extensively investigated, as for example
in \cite{Kausch_1,Kausch_2}. Let us also recall that for $S=2$,
the bosonic base of \CPone\ is a 2-sphere. The superspace \CPone
$=S^2\times \mathbb{C}^{2}$ is a bundle with fermionic complex
2-dimensional fibers. As for their bosonic cousins, the second
homology group $H_2(\CPn) = \mathbb{Z}$ of complex projective
superspaces is non-trivial. Consequently, \CPn\ supports line
bundles whose second Chern-class is characterized by an integer $M
\in \mathbb{Z}$. In the case of \CPone, the expression for the
corresponding connection one-form is well known from the theory of
Dirac monopoles. We shall often refer to the integer $M$ as the
{\em monopole number}.
\medskip

The construction of the sigma model on \CPn\ can be inferred from
the geometric construction we outlined above. The model involves a
field multiplet $Z_\alpha=Z_\alpha(z,\bar z)$ with $S$ bosonic
components $Z_\alpha = z_\alpha, \alpha =1, \dots, S,$ and the
same number of fermionic fields $Z_\alpha = \xi_{\alpha - S},
\alpha = S+1, \dots ,2S$.
To distinguish between bosons and fermions we introduce from now on a grading function $|\cdot|$, which is 0 when evaluated on the labels
of bosonic and 1 on the labels of fermionic quantities. 
In addition we also need a non-dynamical
U(1) gauge field $a$. With this field content, the action takes
the form
\begin{equation} \label{action}
S \ = \ \frac{1}{2g_\sigma^2} \int d^2z  (\pl_\mu - ia_\mu)
Z_\alpha^\dagger (\partial_\mu + i a_\mu)Z_\alpha -
\frac{i\theta}{2\pi} \int d^2z\epsilon^{\mu\nu} \pl_\mu a_\nu
\end{equation}
and the fields $Z_\alpha$ are subject to the constraint
$Z_\alpha^\dagger Z_\alpha = 1$. \footnote{Note that we eliminated
the radius $R$ of the complex projective space in favor of a
coupling $g_\sigma^{-2}$ entering the action in front of the
metric. Equivalently, we can set $g_\sigma^2=1$ and work with a
radius parameter $R$ appearing in the modified constraint
$Z_\alpha^\dagger Z_\alpha = 4R^2$.} The integration over the
abelian gauge field can be performed explicitly and it leads to
the replacement
\begin{equation}
a_\mu \ = \ \frac{i}{2} \left[ Z_\alpha^\dagger \pl_\mu Z_\alpha -
(\pl_\mu Z_\alpha^\dagger) Z_\alpha\right] \ \ .
\end{equation}
The term multiplied by $\theta$ does not contribute to the
equations of motion for $a_\mu$. As its bosonic counterpart, the
\CPn\ sigma model on a closed surface possesses instanton
solutions. The corresponding instanton number is computed by the
term that multiplies the parameter $\theta$. Since it is integer
valued, the parameter $\theta = \theta + 2\pi$ can be considered 
periodic as long as the world-sheet has no boundary.
\smallskip

In order to pass to our second formulation of the \CPn\ model we
employ the gauge freedom to solve the constraint $Z_\alpha^\dagger
Z_\alpha = 1$ as follows
\begin{equation}\label{eq:pr_par}
 Z_1 =Z^\dagger_1\ = \ \frac{1}{\sqrt{1+ w^\dagger \cdot w}}, \quad
 Z_{i+1} \ = \ \frac{w^i}{\sqrt{1+ w^\dagger \cdot w}},\quad
 Z^\dagger_{i+1}\  = \ \frac{w^{\bar{\imath}}}{\sqrt{1+ w^\dagger \cdot
 w}}\ .
\end{equation}
Thereby we have parametrized the target space \CPn\ through a set
of $S-1$ complex bosonic components $w_1,\dots, w_{S-1}$ and a set
of $S$ complex fermionic ones $w_{S},\dots, w_{2S-1}$. Plugging
this parametrization~\eqref{eq:pr_par} back into the
action~\eqref{action} we obtain an unconstrained reformulation of
the \CPn\ sigma model
\begin{equation}\label{eq:act2}
 S \ =\  \frac{1}{2g_\sigma^2}\int d^2z\, g_{i\bar{\jmath}}
 \partial_\mu w^{\bar{\jmath}} \partial_\mu w^i +
 \frac{i\theta}{2\pi} \int d^2z\,
 \epsilon^{\mu\nu} i g_{i\bar{\jmath}}\partial_\nu
 w^{\bar{\jmath}} \partial_\mu w^i,
\end{equation}
where $g_{i\bar{\jmath}}$ is the canonical Fubini-Study metric
on \CPn
\begin{equation}\label{eq:fs_metric_on_cp_blabla}
 g_{i\bar{\jmath}} \ =\  \frac{\delta_{ij}}{1+w^\dagger\cdot w} -
 \frac{(-1)^{|j|}w^{\bar{\imath}}w^j}{(1+w^\dagger\cdot w)^2}\ .
\end{equation}
The disadvantage of this reformulation is the non-linear action of
the \Un\ supergroup on the projective coordinates $w,\bar{w}$. Let
us recall in passing that the Fubini-Study metric on \CPn\
determines the following K\"ahler two-form
\begin{equation}\label{eq:Kahler_form}
 \omega \ =\  d^2z\,\epsilon^{\mu\nu} i g_{i\bar{\jmath}}
 \partial_\nu w^{\bar{\jmath}} \partial_\mu w^i \ =\
 -i g_{i\bar{\jmath}} dw^{\bar{\jmath}} \wedge dw^i\ .
\end{equation}
The K\"ahler form is properly normalized and generates the
second integral cohomology group of \CPn, that is
\begin{equation}
 \int \frac{\omega}{2\pi}\  = \ 1 \ .
\end{equation}
It follows, as stated before, that our bulk model is not affected if we shift
$\theta$ by integer multiples of $2\pi$, i.e.\ we can restrict the
parameter $\theta$ to the interval $\theta \in [-\pi, \pi[$.

\subsection{Action of the boundary model}

We are now going to discuss \Un\ symmetric boundary conditions of
the \CPn\ model. For readers used to the string theoretic concept
of branes and the geometric classification of boundary conditions,
the final outcome is not surprising. Note that \CPn\ admits a
natural left action of \Un. Since \CPn\ is homogeneous under this
action, any \Un\ symmetric brane must be volume filling. But branes
are not simply (sub-)manifolds in target space. They also carry a
bundle ${\cal L}$ with connection $A$. In the case at hand, there
is an infinite family of complex line bundles ${\cal L}_M$ on \CPn\
which are parametrized by the integer $M \in \mathbb{Z}$. To ensure
\Un\ invariance, the connection $A_M$ must have constant curvature
$\Omega_M$. Consequently, its  curvature is proportional to the
K\"ahler form $\omega$, i.e.\ $\Omega_M \sim M \omega$. We shall
now see how these geometric insights manifest themselves in the
world-sheet description. Our presentation will not make any more
reference to string theoretic notions.

We want to consider the \CPn\ sigma model on a world-sheet
$\Sigma$ with boundary. The choice we have in mind is a strip
$\Sigma = [0,\pi] \times \mathbb{R}$ or, equivalently,  the upper
half of the complex plane $z = x+iy, y > 0$. We are looking for
boundary conditions which arise from adding boundary terms of
the form
\begin{equation}\label{eq:bound_act}
 S_b \ = \ \int_{-\infty}^0 dx\, \left(A^L_i(w,\bar{w}) \partial_x w^i +
 A^L_{\bar{\imath}}(w,\bar{w}) \partial_x w^{\bar{\imath}}\right) +
    \int_0^\infty dx\, \left(\, L \ \leftrightarrow \ R \, \right)
\end{equation}
to the action~(\ref{action}, \ref{eq:act2}). Here, $A^{L(R)} =
A^{L(R)}_idw^i+A^{L(R)}_{\bar{\imath}}dw^{\bar{\imath}}$ are one-forms which
are at least locally defined on \CPn. When we map the half plane
back to the strip, points with $z = x>0$ are mapped to the right
boundary while those with $z=x<0$ end up on the left side. To find
consistent boundary conditions we require invariance of the total
action $S_t=S+S_b$ with respect to arbitrary variations $\delta
w^i(z,\bar{z})$. It follows that
%
%
%
\begin{align}\label{eq:bc}
 g_{i\bar{\jmath}}\left( \frac{1}{2g_\sigma^2}\partial_y  +
 \frac{\theta }{2\pi}\partial_x\right)w^{\bar{\jmath}} &=\
 2\Omega^{L(R)}_{ij} \partial_x w^j+ 2
 \Omega^{L(R)}_{i\bar{\jmath}}\partial_x w^{\bar{\jmath}}\\[2mm] \notag
 g_{\bar{\imath}j}\left( \frac{1}{2g_\sigma^2} \partial_y  -
 \frac{\theta }{2\pi}\partial_x\right)w^j &=\
 2\Omega^{L(R)}_{\bar{\imath}\bar{\jmath}} \partial_x w^{\bar{\jmath}}+
 2\Omega^{L(R)}_{\bar{\imath}j}\partial_x w^j,
\end{align}
where $z=\bar{z}<0 \, (>0)$ and $\Omega^{L(R)}$ is the curvature
2-form of the connection $A^{L(R)}$. It is globally defined on
\CPn\ through
\begin{equation}
  \Omega^{L(R)} = dA^{L(R)} \ =\  -\Omega^{L(R)}_{ij}dw^j\wedge dw^i
  -2\Omega^{L(R)}_{i\bar{\jmath}} dw^{\bar{\jmath}}\wedge dw^i-
  \Omega^{L(R)}_{\bar{\imath}\bar{\jmath}}dw^{\bar{\jmath}}\wedge
dw^{\bar{\imath}}.
\end{equation}
Before imposing the conditions of  \Un\ symmetry, we note that our
boundary conditions~\eqref{eq:bc} should preserve the complex
structure of the \CPn\  supermanifold. Consequently, the complex
conjugate of the first equation in~\eqref{eq:bc} must yield the
second equation without any additional constraint. While applying
this constraint one must take into account the reality condition
of a scalar field in Euclidean space-time
\begin{equation}
  w^i(z,\bar{z})^* \ =\  w^{\bar{\imath}}(1/z^*,1/\bar{z}^*).
\end{equation}
Thus, we conclude that the two equations in~\eqref{eq:bc} are
compatible if and only if $\Omega^{L(R)}$ is \emph{imaginary}.

Boundary conditions~\eqref{eq:bc} are said to preserve the global
\Un\ symmetry if they are invariant with respect to an
infinitesimal action of the supergroup. To give a precise meaning
to this statement, let us note that the set of
eqs.~\eqref{eq:bc} can be interpreted as the vanishing
condition of some real vector field $L$ on the left boundary, with
components
\begin{align}\label{eq:bc_vf}
  L^i &=\  \left(\frac{1}{2g_\sigma^2}\partial_y  -
  \frac{\theta}{2\pi}\partial_x\right) w^i
  -2g^{i\bar{k}}\Omega^L_{\bar{k}\bar{\jmath}} \partial_x w^{\bar{\jmath}}
  -2g^{i\bar{k}}\Omega^L_{\bar{k}j} \partial_x w^j,\\[2mm]
  L^{\bar{\imath}} &=\  \left(\frac{1}{2g_\sigma^2}\partial_y  +
  \frac{\theta}{2\pi}\partial_x\right)w^{\bar{\imath}}
  -2 g^{\bar{\imath}k}\Omega^L_{kj} \partial_x w^{j}
  -2 g^{\bar{\imath}k}\Omega^L_{k\bar{\jmath}} \partial_x w^{\bar{\jmath}}
\end{align}
and a similar expression for another real vector field $R$ on the
right boundary. The global \Un\ invariance of the boundary
conditions~\eqref{eq:bc} is then equivalent to the invariance of
the vector field $L,R$ with respect to the infinitesimal action of
the \un\ Lie superalgebra. In other words, the Lie derivative of
the vector field~\eqref{eq:bc_vf} with respect to any \un\ Killing
vector must vanish. As world-sheet translations and global
symmetry transformations commute, it follows that the 2-forms
$\Omega^{L(R)}$ must be invariant. On the other hand, on any
irreducible complex symmetric superspace, there is only one
invariant closed 2-form, namely the K\"ahler form $\omega$. Hence,
invariance of the boundary conditions with respect to the global
symmetry requires that
\begin{equation}\label{eq:curv_conn}
 \Omega^L \ = \  i M \omega \ \quad \ \quad , \quad \quad
 \Omega^R \ = \ i N \omega
\end{equation}
where $\omega$ is the K\"ahler form~\eqref{eq:Kahler_form}. In the
classical theory, the $M,N$ can assume any real value. For the
associated path integral to be well defined, however, they must be
integers. Even though this sections deals with the classical
action, we shall assume $M,N \in \mathbb{Z}$ from now on. For
later use it is convenient to re-write $L$ and $R$ in an index
free notation,
\begin{equation} \label{LRdef}
  L \ = \ \frac{1}{2g_\sigma^2}\partial_y + i J
  \left(\frac{\theta}{2\pi}+M\right) \partial_x\quad ,\quad
  R \ = \ \frac{1}{2g_\sigma^2}\partial_y + i J
  \left(\frac{\theta}{2\pi}+N\right)\partial_x \ .
\end{equation}
Here we have introduced the (globally well defined) complex
structure $J$ on the tangent space of \CPn. The components $L^i =
dw^i(L)$ and $L^{\bar \imath}=dw^{\bar \imath}(L)$ are recovered from $L$
with the help of the canonical basis $dw^i$ and $dw^{\bar \imath}$ in the
cotangent space. Note that $L$ and $R$ only contain a specific
combination $\Theta_1 = 2M+\theta/\pi$ and, respectively,
$\Theta_2 = 2N+\theta/\pi$. We conclude that the
periodic variable $\theta$ of the bulk theory gets promoted to a
real valued variable $\Theta$ in the boundary problem \cite{Xiong}. 
In the limit $g_\sigma \rightarrow 0$, the value of $\Theta$ is
irrelevant. In other words, the boundary conditions are purely
Neumann when we approach infinite radius.

Before we close this section, let us briefly write the boundary
conditions in terms of the manifestly \Un\ covariant formulation
\eqref{action} of our theory. In this case, the variations of the
basic fields $Z_\alpha$ must be consistent with the  constraint
eq.\ \eqref{SSconstraint},
$$ (\delta Z^\dagger_\alpha) Z_\alpha + Z_\alpha^\dagger \delta
Z_\alpha = 0 \  \ . $$
In order for the boundary contributions to the variation of the
action to vanish, we must impose the usual twisted Neumann
boundary conditions of the type
\begin{equation} \label{glue}\begin{split}
(\partial_y + i a_y) Z_\alpha & = \ \ \Theta_1 g_\sigma^2
(\partial_x + i a_x) Z_\alpha\  ,\\[2mm]
(\partial_y - i a_y) Z^\dagger_\alpha & =\  - \Theta_1
g_\sigma^2 (\partial_x - ia_x) Z_\alpha^\dagger
\end{split}
\end{equation}
for $z = \bar z < 0$ and a similar condition with $M$ replaced by
$N$, i.e. $\Theta_1$ replaced by $\Theta_2$, along the right half $z = \bar z > 0$ of the boundary. The
parameters $\Theta_1 = 2M + \theta/\pi$ and $\Theta_2=2N + \theta/\pi$ are the same combination of
the $\theta$ angle in the bulk and the monopole numbers $M,N$ that
appeared in eq.\ \eqref{LRdef}.

So far we have only discussed the classical theory. Understanding
the detailed properties of the associated quantum field theories
is the main aim of the following sections. For the time being let
us just mention that the non-linear sigma models on \CPn\ have
been argued to possess vanishing $\beta$ function
\cite{Read:2001pz}. This means that
they give rise to conformal quantum field theories for any choice
of the two couplings $g_\sigma$ and $\theta$. The central charge
of these models must agree with the central charge of the free
field theory at $g_\sigma \sim 0$, i.e.\ all models of this type
have $c= -2$.

\section{Spectrum of the non-interacting sigma model}

Our discussion of the quantum field theory will begin with the
limiting case $g_\sigma =0$ in which all the interactions are
turned off. To keep things explicit, we will restrict to the first
non-trivial case with $S=2$. Most of what we are about to describe
generalizes quite easily to higher dimensional projective
superspaces. We want to investigate the spectrum of the \CPone\
model on the strip (or half-plane) with twisted Neumann boundary
conditions imposed along the boundary. In more stringy terms this
corresponds to considering volume filling branes which wrap the
2-sphere of \CPone. In a first step we shall analyze the spectrum
in the particle limit. Then, in the second step, we include
derivative fields and construct a partition function for the
theory in the limit of vanishing coupling $g_\sigma$.

\subsection{Spectrum for a particle moving on \CPone}

The semiclassical or minisuperspace approximation amounts to
considering the string as a point like object, that is to
neglecting the $\sigma$ dependence of the fields
$w^i(\tau,\sigma)$. Thereby, we reduce the field theory to a point
particle problem. We shall discuss the quantization of this
system in two different ways. In the first description we use the
gauge fixed formulation of the theory in terms of variables $w^i,
w^{\bar \imath}$. The spectrum of the associated Hamiltonian is known
from \cite{kuwabara}. Our second approach employs the U(2$|$2)
covariant formulation. Its results agree with the first treatment,
but the U(2$|$2) covariant construction is extended more easily to
the full field theory.

So, let us start from the action \eqref{eq:act2} and set all
$\sigma$-derivatives to zero.
Integrating out the transverse coordinate $\sigma$ of the strip $\Sigma=[0,\pi]\times\mathbb{R}$ we get the following particle theory
\begin{equation}
  S \ = \ \int_{-\infty}^\infty d\tau\, \left(\frac{\pi}{2g_\sigma^2}
  g_{i\bar{\jmath}}\dot{w}^{\bar{\jmath}}\dot{w}^i+ A_i \dot{w}^i +
  A_{\bar{\imath}}   \dot{w}^{\bar{\imath}} \right)\ ,
\end{equation}
where locally the connection one-form $A$ is the difference of the
two one-forms $A^R$ and $A^L$, i.e.\
\begin{equation}\label{eq:tachyonic_bundle}
  A \ = \ A^R - A^L\ \ .
\end{equation}
The classical Hamiltonian of this quantum mechanical system takes
the following simple form
\begin{equation}\label{eq:cl_ham}
  H \ =\  -\frac{2g_\sigma^2}{\pi} g^{i\bar{\jmath}}(\Pi_i -
  A_i)(\Pi_{\bar{\jmath}}-A_{\bar{\jmath}})\ \ ,
\end{equation}
where the canonical momenta are given by
\begin{equation}
  \Pi_i \ =\
  \frac{\pi}{2g_\sigma^2}g_{i\bar{\jmath}}\dot{w}^{\bar{\jmath}}+A_i\ ,\quad
  \Pi_{\bar{\imath}} \ =\
  \frac{\pi}{2g_\sigma^2}g_{\bar{\imath}j}\dot{w}^{j}+A_{\bar{\imath}}\ \ .
\end{equation}
We can now pass to the quantum theory through the usual canonical
quantization, i.e.\ by replacing Poisson brackets with
commutators,
\begin{equation}
  [w^i , \Pi_j] \ = \ [w^{\bar{\imath}},\Pi_{\bar{\jmath}}] \ =\ \delta^i_j\
  \ .
\end{equation}
Note that the factor $i$ of the usual commutation relations
$[x^i,p_j]=i\delta^i_j$ is missing because we are formulating the
theory in Euclidean time $\tau = it$. For the quantization
procedure to make sense, the one-form $A$ must be a connection on
a complex line bundle over \CPone, see \cite{greub74}. This
furnishes a quantization condition for the curvature of the
connection,
\begin{equation}
  d A \ = \ -il \omega
\end{equation}
with $l$ any integer and $\omega$ the K\"ahler form on \CPone. The
space of sections of such bundles may be realized explicitly as
equivariant functions $f(w,\bar{w})$ on \CPone\ with the property
\begin{equation}
  f(e^{i \alpha} w, e^{-i \alpha} \bar{w}) =
  e^{il\alpha}f(w,\bar{w})\ .
\end{equation}
Taking into account~\eqref{eq:tachyonic_bundle} we get the
condition that $l = M-N$ must necessarily be an integer. Hence, if
we admit e.g. $A^L=0$ as a possible boundary conditions, mutual
consistency requires $N \in \mathbb{Z}$. The quantized form of the
classical Hamiltonian~\eqref{eq:cl_ham} is, up to a numerical
prefactor, the Bochner-Laplacian
$\Delta^{(l)}_{\CPn}$ on the complex line bundle
over \CPone\ with monopole charge $l \in \mathbb{Z}$
\begin{equation}
  \hat{H}^{(l)} \ = \ -\frac{g_\sigma^2}{\pi}
  \Delta^{(l)}_{\CPn}\ .
\end{equation}
The eigenvalues of the Bochner-Laplacian on
\CPn\ where studied in~\cite{kuwabara}. For the Hamiltonian we
obtain
\begin{equation} \label{eq:spectrum}
  h_{l}(k) \ = \ \frac{g_\sigma^2}{\pi}
  \left(2k^2 + (2k+|l|)(|l|-1) - l^2\right)\quad \mbox{ for }
  \ k \ = \ 0,1,2, \dots
\end{equation}
From the spectrum we can read off which \Utwo\ multiplets are
realized as sections of monopole bundles on \CPone. We will list
the corresponding  representations of U(2$|$2) a bit later at the
end of our second construction of the spectrum.

Let us now see how to reproduce the spectrum of the particle
theory within the U(2$|$2) covariant formulation. As before, we
depart from the space $\mathbb{C}^{2|2}$ with coordinates
$Z=(z_1,z_2,\xi_1,\xi_2)$. The 4-tuple $Z$ transforms in the
fundamental representation $V$ of \utwo. On the projective
superspace \CPone, the multiplet $Z$ and its conjugate $Z^\dagger$
obey the following constraint
\beq \label{constraint} Z^{\dagger}\cdot Z\ =\ 1\qquad . \eeq
Note that $Z^{\dagger}$ transforms in the dual fundamental
representation $Z^{\dagger}\in V^\star$ so that the equation
\eqref{constraint} is consistent with the \utwo\ symmetry.
Consequently, if we quotient the space of functions on
$\mathbb{C}^{2|2}$ by the ideal that is generated from $Z^\dagger
\cdot Z - 1$, we end up with some non-trivial \utwo\ module ${\cal
B}$. The center of \utwo\ acts on ${\cal B}$ through the phase
rotations \eqref{phrot}, thereby defining a decomposition ${\cal
B} = \bigoplus_l {\cal B}_l$ where ${\cal B}_l \subset{\cal B}$
consists of elements $f\in {\cal B}$ such that $f \rightarrow \exp
(il\varpi) f$ under the map \eqref{phrot}. The spaces ${\cal B}_l$
contain precisely all sections of the complex line bundle with monopole
number $l$.

We want to determine the partition function of the particle limit,
i.e.\ a function that counts sections in the monopole line
bundles, or, equivalently, elements in the \utwo\ module ${\cal B}_l$.
Before we construct this counting function, let us introduce the
following basis in the 4-dimensional Cartan subalgebra,
\beq \label{Cartanutwo}
J_x=\frac12 \, \left(\begin{array}{c|c} \sigma_3 & 0\\
\hline 0 &
0\end{array}\right) \quad J_y =\frac12\, \left(\begin{array}{c|c} 0 & 0 \\
\hline 0 & \sigma_3
\end{array}\right)\quad J_z=\frac12\, \left(\begin{array}{c|c} I_2 & 0
\\\hline  0 & -I_2\end{array}\right)\quad J_u=\frac{I_4}{2} \ . \eeq
Here $I_n$ is the $n$-dimensional identity and $\sigma_3$ the Pauli
matrix $\sigma_3 = $diag$(1,-1)$. The partition function reads
\beq Z^{(0)}_{M,N}(x,y,z)\ = \ \tr_{{\cal B}_l}(x^{J_x} y^{J_y}
z^{J_z}) = \lim_{t\rightarrow 1}\oint_{|u|=1}\!\!\!\! du\
\frac{1-t^2}{u^{l/2+1}}\!\!\prod_{\alpha,\beta = \pm \frac12}
\frac{(1+y^{\alpha} z^{-\beta} u^{\beta} t
)}{(1-x^{\alpha}z^{\beta}u^{\beta}t)} \ , \eeq
where $l=M-N$ is the difference of the monopole numbers, as
before. The trace is taken over all sections of line bundles on
\CPone\ and the integral over $u$ is to be understood in the
formal sense, i.e. as a projector. The limit $t \rightarrow 1$
implements the constraint \eqref{constraint} (see
\cite{Mitev:2008yt} for details) while the integral over the
variable $u$ selects those states that stretch between two line
bundle with monopole number $N$ and $M$, respectively. Of course,
states within $Z^{(0)}_{M,N}$ still carry a $J_u$ charge. It takes
the constant value $J_u = l/2$.

Our aim now is to decompose the partition function of the particle
theory into characters of the symmetry \utwo. In a first step we
expand $Z^{(0)}$ into characters of 8-dimensional bosonic
subalgebra $\utwo_{\overline 0} \cong \sltwo \oplus \sltwo \oplus
\uone \oplus \uone$. The latter are given by
\beq\label{bchar}  \chi^B_{\left(j_1,j_2,a,b\right)}(x,y,z,u)\ =\
\chi_{j_1}(x)\, \chi_{j_2}(y)\,  z^a\,  u^b\ , \eeq
where $j_1,j_2\in \mathbb{N}/2$ and $a,b\in \mathbb{C}$. It is
rather straightforward to compute the corresponding branching
functions and we shall not spell out the results of this
intermediate step here. A similar computation in the case of
supersphere sigma models can be found in \cite{Mitev:2008yt}. The
next step then is to combine the characters of the bosonic
subalgebra into characters of \utwo. Two types of characters turn
out to appear. The generic ones are the characters of Kac-modules,
i.e.\ of irreducible and degenerate long multiplets. Their relation to
characters of the bosonic subalgebra is given by
\beq \label{Kmod} \chi^K_{\left[j_1,j_2,a,b\right]} =
\chi^B_{\left(j_1,j_2,a,
b\right)}\left(1+z^{-1}\chi_{\left(\frac{1}{2},\frac{1}{2}\right)}+z^{-2}
\left(\chi_{(1,0)}+\chi_{(0,1)}\right)+z^{-3}\chi_{\left(\frac{1}{2},
\frac{1}{2} \right)}+z^{-4}\right). \eeq
Here and in the following we abbreviate the products
$\chi_{j_1}(x)\chi_{j_2}(y)$ of \sltwo-characters as
$\chi_{\left(j_1,j_2\right)}$. In this expression, the first
factor is associated with the bosonic multiplet of ground states
while the expression within brackets arises from the four
fermionic lowering operators in a Kac-module of \utwo. In addition
to the Kac-modules, we also need formulas for characters of some special
atypical irreducibles. According to \cite{ZhangGould}, the
characters of these atypicals are given by
\beqa \label{Amod}
\chi_{\left[\frac{l}{2},0,\frac{l}{2},\frac{l}{2}\right]}&\
=\ &\chi^B_{\left(\frac{l}{2}, 0,\frac{l}{2}, \frac{l}{2}\right) } +\chi^B_ {
\left(\frac{l-1}{2},\frac{1}{2},\frac{l-2}{2},\frac{l}{2}\right)}
+\chi^B_{ \left(\frac{l-2}{2}, 0 , \frac{l-4}{2} ,
\frac{l}{2}\right)}\nonumber\\[2mm]
\chi_{\left[\frac{l-2}{2},0,\frac{4-l}{2},-\frac{l}{2}\right]}&\
=\ &\chi^B_{
\left(\frac{l}{2}, 0 , -\frac{l}{2} ,
-\frac{l}{2}\right)}+\chi^B_{\left(\frac{l-1}{2},\frac{1}{2},-\frac{2+l}{2},
-\frac {l}{2}\right)}
+\chi^B_{\left(\frac{l-2}{2}, 0,\frac{4-l}{2}, -\frac{l}{2}\right) }
\nonumber\\[2mm]
\chi_{\left[\frac{l+1}{2},\frac{1}{2},\frac{l+2}{2},\frac{l}{2}\right]}&\
=\ &\chi^B_{\left(\frac{l+1}{2}, \frac{1}{2},\frac{l+2}{2},
\frac{l}{2}\right) } +\chi^B_ {
\left(\frac{l}{2},0,\frac{l}{2},\frac{l}{2}\right)} +\chi^B_{
\left(\frac{l+2}{2}, 0 , \frac{l}{2} ,
\frac{l}{2}\right)}+\chi^B_{ \left(\frac{l}{2}, 1 , \frac{l}{2} ,
\frac{l}{2}\right)}\\[2mm]
&&+\, \chi^B_{ \left(\frac{l+1}{2}, \frac{1}{2} , \frac{l-2}{2} ,
\frac{l}{2}\right)}+\chi^B_{ \left(\frac{l-1}{2}, \frac{1}{2} ,
\frac{l-2}{2} , \frac{l}{2}\right)}+\chi^B_{ \left(\frac{l}{2}, 0
, \frac{l-4}{2} ,
\frac{l}{2}\right)}\nonumber\\[2mm]
\chi_{\left[\frac{l}{2},0,\frac{4-l}{2},-\frac{l}{2}\right]}&\ =\
& \chi^B_ {
\left(\frac{l+1}{2},\frac{1}{2},\frac{2-l}{2},-\frac{l}{2}\right)}+\chi^B_{
\left(\frac{l}{2}, 0 , -\frac{l}{2} , -\frac{l}{2}\right)}
+\chi^B_{ \left(\frac{l+2}{2}, 0 , -\frac{l}{2} ,
-\frac{l}{2}\right)}+\chi^B_{ \left(\frac{l}{2}, 1 , -\frac{l}{2}
, -\frac{l}{2}\right)}\nonumber\\[2mm] &&+\chi^B_{ \left(\frac{l+1}{2},
\frac{1}{2} , \frac{2-l}{2} , -\frac{l}{2}\right)}+\chi^B_{
\left(\frac{l-1}{2}, \frac{1}{2} , \frac{2-l}{2} ,
-\frac{l}{2}\right)}+\chi^B_{\left(\frac{l}{2}, 0,\frac{4-l}{2},
-\frac{l}{2}\right) } \nonumber \ , \eeqa
where $l\geq 0$ and the value $l = -1$ is admitted only in the
third equation. It is understood that a bosonic character is to be omitted on
the right hand side if one of its first two labels is negative. We also
note that $\left[\frac{1}{2},0,\frac{1}{2},\frac{1}{2}\right]$ and
$\left[0,\frac{1}{2},\frac{1}{2},-\frac{1}{2}\right]$ are
associated with the fundamental representation and its dual. The
formulas \eqref{Kmod} and \eqref{Amod} are the only ones we need
in order to obtain the expansion of the partition function in
terms of characters of \utwo
\beqa Z^{(0)}_{M,N}(x,y,z) & = &
 \left(1+\delta_{l,0}\right)\, \chi_{\left[\frac{|l|}{
2 } , 0,\frac{l+2-2\text{sgn}(l)}{2},\frac{l}{2}\right]} +
\label{Z0part}\\[2mm]
& & + \, \chi_{\left[ \left|\frac{l}{2}+\frac { 3 }
{4}\right|-\frac{1}{4} ,
\frac{1+\text{sgn}(l+1)}{4},\frac{l+3-\text{sgn}(l+1)}{2},\frac{l}{2}\right]}
+ \sum_{k=2}^{\infty} \chi^K_ { \left [ k-1+\frac { |l| } { 2},
0,\frac{|l|}{2}+2,\frac{l}{2}\right]}  \ , \nonumber\eeqa
where $l = M-N$ and $\text{sgn}(x)=1$ if $x\geq 0$,
$\text{sgn}(x)=  -1$ otherwise. The first two summands in this
formula involve characters of irreducible atypicals while all
remaining ones are associated with full Kac-modules. In the
special case that $l=M-N =0$, the partition function is counting 
functions on \CPone. Note that the last label $b$ of all
representations becomes trivial for $l=0$ meaning that we are
dealing with representations $[j_1,j_2]_p = [j_1,j_2,p,0]$ of the
quotient \putwo. If we restrict further to the subalgebra
$\psutwo$ we may combine the atypical characters into the
character of a single (atypical) Kac-module $[0,0]$. Consequently,
the decomposition contains contributions from $[k,0]$ with $k =
0,1,2, \dots$ These are the characters\footnote{See
\cite{Gotz:2005ka} for more details on the representation theory
of $\psutwo$.} of the \psutwo\ supermultiplets which are generated
from spherical harmonics on the bosonic base of \CPone. For values
$l = M-N > 0$, the lowest value of $j_1$ is $j_1 = |l|/2$. Such a
cutoff is a well known feature of sections in monopole bundles.

The result \eqref{Z0part} agrees with our earlier description of
the spectrum \eqref{eq:spectrum}. To relate the two findings we
note that in a representation $[j_1,j_2,a,b]$ of \utwo\ the
quadratic Casimir elements take the value
\begin{equation} \label{eq:Casimiralpha}
    \Cas_\alpha(\Lambda) \ = \ 2[j_1(j_1+1)- j_2(j_2+1) + b(a-2)]
    - 4\alpha b^2.
\end{equation}
Since \Utwo\ is not semisimple, there exists a one-parameter
family of such Casimir elements. It is parametrized by the
coefficient $-\alpha$ of $E^2$ where $E$ denotes the central
element of \utwo. More details can be found in Appendix
\ref{sec:most_gen_2_cas}. Plugging in the labels of
representations from eq.\ \eqref{Z0part} one recovers the spectrum
\eqref{eq:spectrum} of the Bochner-Laplacian, provided the parameter $\alpha$
in the Casimir element is set to $\alpha =1$ (see Appendix \ref{sec:Laplacian}
for details). This concludes our discussion of the particle limit.

\subsection{Partition function at infinite radius}

The partition function of the boundary conformal field theory in
the limit of vanishing target space curvature can be constructed
by extending our discussion of the particle limit to incorporate
derivatives along the boundary. The main formula is
\beqa \label{completespectrum2} Z_{M,N}(x,y,z;q)&=& \phi(q)\,
q^{\frac{1}{12}}\oint_{|u|=1}\frac{du}{u^{l/2+1}}\, \phi(q) \,
\lim_{t\rightarrow 1}(1-t^2) \, \times \nonumber\\[2mm]
& &  \hspace*{-15mm} \times \
\prod_{\alpha,\beta=\pm\frac{1}{2}}\frac{1+y^{\alpha}
\left(zu^{-1}\right)^{\beta} t}{1-x^{\alpha}
\left(zu\right)^{\beta} t}
\prod_{n=1}^{\infty}\prod_{\alpha,\beta=\pm\frac{1}{2}}
\frac{1+y^{\alpha} \left(zu^{-1}\right)^{\beta}q^n}{1-x^{\alpha}
\left(zu\right)^{\beta}q^n}\ . \eeqa
As in the particle model, the limit $t \rightarrow 0$ is used to
implement the constraint \eqref{constraint} on the zero modes of
fields. But in the field theory, there is an infinite tower of
field identifications that follow by taking derivatives. Once the
constraint on zero modes has been taking into account, the net
effect of the remaining field theoretic identifications is to
remove local fluctuations in one bosonic direction from the state
space. This is achieved my multiplying the counting function with
the Euler function $\phi(q)$. A more thorough mathematical derivation
of this argument may be found in \cite{Mitev:2008yt}. The line
integral over $u$ projects onto fields that possess the same
behavior under the global phase rotations \eqref{phrot} of fields.
In the field theory, however, local phase rotations are gauged by
the non-dynamical gauge field $a$. Thereby, we remove fluctuations
into a second bosonic direction. On the level of our partition
function the double counting of fields which are related by local
gauge transformations is avoided by another multiplication with the
 Euler function $\phi(q)$.

Now that we understood our basic expression from the partition
function of the model, let us decompose the field theory spectrum
into representations of the global symmetry \utwo. As in the
particle limit, we expand into bosonic characters first,
$$ Z_{M,N}(x,y,z;q) \ = \ \sum \chi^B_{(j_1,j_2,a,b)}(x,y,z) \,
\psi^{B}_{(j_1,j_2,a,b)}(q) \ \ ,
$$
where $b=l/2$ and the sum runs over all $j_1,j_2\in
\frac{\mathbb{N}}{2}$, $a\in \frac{\mathbb{Z}}{2}$ for which $a+b\equiv 2j_1
\text{ mod } 2$ and  $a-b\equiv 2j_2 \text{ mod
} 2$. The characters
$\chi$ of the even part \utwo$_{\overline 0}$ were
displayed in equation \eqref{bchar} above. For the associated
branching functions $\psi^B$ one finds
\beqa
\psi^{B}_{\left(j_1,j_2,a,b\right)}(q)&=&\frac{q^{\frac{1}{12}}}{\phi(q)^4}
\left(q^{-j_2}-q^{3j_2+2}+\left(q^{\frac{a-b}{2}}+
q^{\frac{b-a}{2}}\right)\left(1-q^{2j_2+1}\right)\right)\, \times
\nonumber\\[2mm]
&&\hspace*{-2.5cm}
\times\, q^{j_2^2+\left(\frac{a-b}
{2}\right)^2} \sum_{\substack{l=\left|\frac{a+b}{2}\right|\\j_1+l\in
\mathbb{N}}}\sum_{m,n=1}^{\infty}(-1)^{m+n}\frac{
\left(1-q^
{ m+n }
\right)\left(q^{(m-n)(j_1-l)}-q^{(m-n)(j_1+l+1)}\right)}{q^{-\frac{m(m-1)+n(n-1)
} {2}}}
,\nonumber \eeqa where we require that $a$ and $b$ be such that
\beq a+b\equiv 2j_1 \text{ mod } 2\qquad a-b\equiv 2j_2 \text{ mod
} 2\ .  \eeq
The branching functions for the Kac-modules of the full
superalgebra \utwo\ can be obtained through the following infinite
sums
\beq \label{branchingKac1}
\psi^K_{[j_1,j_2,a,b]}=\sum_{n=0}^{\infty}(-1)^n\sum_{m=0}^{\left[\frac{n}{2}
\right ] }
\sum_{r,s=0}^{n-2m}\psi^B_{\left(j_1+\frac{n}{2}-(m+r),j_2+\frac{n}{2}
-(m+s),a+n,b\right) } \ .
\eeq%
Weights $[j_1,j_2,a,b]$ of \utwo\ are atypical when $b = \pm
(j_1-j_2)$ or $b =  \pm (j_1 + j_2 + 1)$. Whenever the weights are
atypical, our expressions for $\psi^K$ must be summed further to
obtain branching functions of irreducible representations. The
necessary formulas are listed in Appendix \ref{sec:atypicalBF}.
Here, we shall simply display our results in terms of the
branching functions $\psi^K$,
\begin{equation}
Z_{M,N}(x,y,z;q) \ = \ \sum \chi^K_{[j_1,j_2,a,l/2]}(x,y,z) \,
\psi^{K}_{\left[j_1,j_2,a,l/2\right]}(q) \ \ .  \label{ZRinfty}
\end{equation}
The sum runs over all $j_1,j_2\in \frac{\mathbb{N}}{2}$, $a\in
\frac{\mathbb{Z}}{2}$ for which $a+l/2\equiv 2j_1 \text{ mod } 2$
and  $a-l/2\equiv 2j_2 \text{ mod } 2$. For our purposes, the branching
functions $\psi^{K}$ are already good enough, since we are only interested in
the values that the quadratic Casimir takes on the states of our theory and not
in their precise transformation properties which, since indecomposable
representations appear quite naturally, can be very complicated. We recall that
the characters $\chi^K$ of \utwo\ Kac-modules are given by eq.\
\eqref{Kmod}. For typical weights, the functions $\psi^K$ are
proper branching functions with non negative integer coefficients. 

It is very instructive to apply the same combinatorial constructions
to the simpler theory of symplectic fermions, i.e.\ for $S=1$. The
symmetry of this model is described by the superalgebra \uoneone.
We select a particular basis $J_z, J_u$ for the Cartan subalgebra
by fixing the values in the fundamental representation according
to
\beq
J_z\ =\ \frac{1}{2}\left(\begin{array}{c|c}1 & \\ \hline &
-1\end{array}\right)\qquad
J_u\ =\ \frac{1}{2}\left(\begin{array}{c|c}1 & \\ \hline
& 1\end{array}\right)\ .
\eeq
Just as in the case of the \CPone\ model, we construct the partition
function in the limit $R\rightarrow \infty$ by taking tensor products
of the fundamental representation of \uoneone\ and its dual. After that
we apply our constraint and gauge prescription. The partition function
$Z = \sum_l Z_l u^{l/2}$ for all bundles is then given by the formula
\beqa
Z(q|z,u)&=&q^{\frac{1}{12}}\phi(q)^2\lim_{t\rightarrow
1}(1-t^2)\prod_{n=0}^{\infty}\frac{(1+z^{-\frac{1}{2}}u^{\frac{1}{2}}q^n)(1+z^{
\frac { 1
} { 2}}u^{-\frac{1}{2}}
q^n)}{(1-z^{\frac{1}{2}}u^{\frac{1}{2}}q^n)(1-z^{-\frac{1}{2}}u^{-\frac{1}{2}}
q^n)}\nonumber\\[2mm]
&=&\frac{q^{\frac{1}{12}}}{\phi(q)}\sum_{\substack{a,b\in
\mathbb{Z}/2\\a+b\in \mathbb{Z}}} z^{a}\left(1+z^{-1}\right)u^b
q^{\frac{(b-a)(b-a+1)}{2}}\ , \eeqa where in the product of the first line we
are instructed to make the formal substitution  $q^0 \rightarrow t$
before evaluating the limit $t \rightarrow 1$. Since Kac module
characters for \uoneone\ are defined by\footnote{In our notations,
the second label $b$ refers to the value of the central element
$E$ of \uoneone. This differs from the notations that were used
e.g.\ in \cite{Schomerus:2005}.}
$$ \chi^K_{\bracket{a}{b}}\ =\ z^{a}\left(1+z^{-1}\right)u^b\ \ , $$
we obtain the following expression for the branching functions
\beq \psi^K_{\bracket{a}{b}}(q)\ =\
\frac{q^{\frac{1}{12}}}{\phi(q)}q^{\frac{(b-a)(b-a+1) }{2}}\qquad
\mbox{ for } a,b\in \mathbb{Z}/{2}, a+b\in\mathbb{Z}\ . \eeq
The quadratic Casimir takes the value $2b(2a-1)-4b^2$ in the Kac
module labeled by $\bracket{a}{b}$. For a given value of $b = l/2
=(M-N)/2$, there are four states of conformal weight $h=0$ in the
spectrum. More precisely, we find that
$$ Z^{(0)}_{M,N}(z,u)  \ = \ \chi^K_{\bracket{l/2}{l/2}}+
    \chi^K_{\bracket{l/2+1}{l/2}}\ \ , $$
where $l=M-N$. When $l\neq 0$, the two \uoneone\ multiplets that
appear in the decomposition of $Z^{(0)}$ are typical. This changes
only for $l=0$. In that case, the two atypical multiplets
$\bracket{0}{0}$ and $\bracket{1}{0}$ combine into a 4-dimensional
projective indecomposable of \uoneone. Such boundary theories of
the symplectic fermions with four ground states were first
constructed in \cite{Creutzig:2006wk}. Let us also observe that the
number of characters in the decomposition of $Z^{(0)}$ agrees with
the number of atypical characters in the corresponding
decomposition \eqref{Z0part} for the \CPone\ model. This is no
coincidence. In fact, one may show that states of the symplectic
fermion model are associated to atypical multiplets of the sigma model on \CPone.

\section{Sigma model perturbation theory}

Our aim here is to spell out formulas for the boundary partition
function of the \CPone\ model any finite couplings $g_\sigma$ and
$\theta$. In the first subsection we shall briefly sketch how the
background field expansion can be adapted to supersymmetric target
spaces and use this formalism to calculate conformal weights of
boundary fields exactly, to all orders in perturbation theory. As
in the case of superspheres, the shift of the conformal weights
turns out to be given by a particular quadratic Casimir element of
\utwo. The results of the first subsection are then combined with
our expression \eqref{ZRinfty} for the free partition function to
construct the full (perturbative) partition function of the
\CPone\ model with Neumann-type boundary conditions.

\subsection{Background field expansion and 2-point functions}

Let us consider a sigma model on an arbitrary K\"ahler
supermanifold of superdimension $2p|2q$. If we parametrize the
supermanifold through real coordinates $\varphi^i$, its action
takes the following form
\begin{equation}
    S[\varphi] \ = \ \frac{1}{2g_\sigma^2}\int_\Sigma d^2z\,
    \big(\partial_\mu \varphi(z),
    \partial_\mu \varphi(z)\big)_{\varphi(z)} +
    \frac{i\theta}{2\pi}\int_{\varphi(\Sigma)}\omega,
\label{eq:action_sigma}
\end{equation}
where $(X,Y)_\varphi$ denotes the scalar product of two vector
fields $X,Y$ at the point $\varphi$ of the supermanifold and
$\omega$ is the K\"{a}hler form. We assume the latter to be
normalized such that $\int_{\phi(\Sigma)} \omega$ is integer. For
the path integral measure we use
\begin{equation*}
    \mathcal{D}[\varphi] \ = \ \prod_{x\in \Sigma}
d\mu\big(\varphi(z)\big),\qquad
    d\mu(\varphi) = \sqrt{g(\varphi)}\, d\varphi^1\dots d\varphi^{2p+2q}\ .
\end{equation*}
The measure may be regularized by putting the theory on a square
lattice with spacing $a$. To evaluate the scalar product we
introduce a basis
$e_i=\overleftarrow{\tfrac{\partial}{\partial\varphi^i}}$ of right
derivatives. Expanding two vectors $X = e_i X^i$ and $Y= e_i Y^i$,
with respect to this basis, we obtain
\begin{equation}
    (X,Y) \ = \ (-1)^{|i|} X^i g_{ij}Y^j=g_{ij}Y^jX^i\ .
\label{eq:sc_prod_comp_vecs}
\end{equation}
Here, the order of factors does certainly matter. From the
symmetry $(X,Y)=(Y,X)$ of the scalar product in the tangent space
we derive the following symmetry of the metric tensor
\begin{equation*}
 g_{ij} \ = \ (-1)^{|i||j|} g_{ji}\ .
\end{equation*}
We are interested in computing perturbatively the partition
function and the correlation functions by the steepest descent
method around the \emph{constant} classical solution
$\varphi(z,\bar z)=\bar{\varphi}$. For arbitrary Riemannian
manifolds, one can perform the perturbation theory in the
background field method by switching to the geodesic coordinates
as defined in \cite{Boulware82}. When dealing with complex spaces,
however, there exists more appropriate coordinates which keep the
complex structure manifest. Let $w^a$ be a set of holomorphic
coordinates for the K\"ahler supermanifold and choose some point
on it with fixed coordinates $w^a_0$. A set of holomorphic
coordinates $v^a$ for the complex supermanifold $\mathcal{M}$ is
called a \emph{normal system of coordinates} at $w^a_0$ if the
metric $g_{a\bar{b}}(v,\bar{v}|w_0,\bar{w}_0)$ is of the form
\begin{equation}
    g_{a\bar{b}}(v,\bar{v}|w_0,\bar{w}_0) = g_{a\bar{b}}(w_0,\bar{w}_0) +
    \sum_{n=1}^\infty \, c_{a\bar{b}a_1\bar{b}_1\cdots a_n
\bar{b}_n}(w_0,\bar{w}_0)
    \ v^{\bar{b}_n}v^{a_n}\dots v^{\bar{b}_1}v^{a_1}.
\label{eq:norm_coord_compl}
\end{equation}
The holomorphic transition functions $w = c_{w_0}(v)$ between the
set of holomorphic coordinates $w$ and the normal coordinates $v$
at $w_0$ are completely fixed by the required form of the
metric~\eqref{eq:norm_coord_compl}. In fact, one can prove by
induction that the transition functions $c_{w_0}(v)$ must possess
the following power series expansion in $v$
\begin{eqnarray}
    w^\sigma & = & c^{\sigma}_{w_0}(v) \ = \ w^{\sigma}_0 +
    \sum_{n=1}^\infty \frac{1}{n!}\Big(\nabla_{v}^{n-1} v\Big)\Big\vert_{w_0}
(w_0)  \\[2mm]
   &  = &  w^{\sigma}_0 + v^{\sigma}
    -\sum_{n=2}^\infty \frac{1}{n!}\Gamma^{\sigma}_{b_1b_2;b_3;\dots
;b_n}\Big\vert_{w_0}
    v^{b_n}\cdots v^{b_3}v^{b_2}v^{b_1} \ .
\label{eq:hol_trans_func}
\end{eqnarray}
Here, $\nabla$ denotes the covariant derivative on the K\"ahler
manifold. It involves the Christoffel symbols which may be
computed from the metric according to
\begin{equation*}
\Gamma^i_{jk}\ = \ g^{il} \frac{\partial}{\partial w^k} g_{lj}
\ . 
\end{equation*}
In eq.\ \eqref{eq:hol_trans_func} we have expressed the expansion
coefficients through multiple covariant derivatives
$\Gamma^{\sigma}_{b_1b_2;\dots;b_n}$ of the Christoffel symbols
$\Gamma^{\sigma}_{b_1b_2}$. When evaluating these derivatives, we
only treat the lower labels $b_i$ as tensor indices, i.e.\ the
covariant derivatives do not act on the label $\sigma$.

In order to actually compute the
metric~\eqref{eq:norm_coord_compl} we use a nice trick. Namely, we
propose to consider some holomorphic mapping $w^a(\zeta)$ from a
compact Riemann surface $\Sigma$, parametrized by the holomorphic
coordinate $\zeta$, to the complex symmetric space that is
parametrized by the  holomorphic coordinates $w^a$. Since the
components of vector fields are known in any frame, the metric in
normal coordinates $v$ at $w_0$ may be derived from the equation
\begin{equation}
    \big(\partial w(\zeta),
    {\bar \partial \bar w(\bar \zeta)}\big)_{w(\zeta)}\  = \
    \big(\partial v(\zeta),
    {\bar \partial \bar v(\bar \zeta)} \big)_{\{v(z),w_0\} }\ .
\label{eq:change_coord_compl}
\end{equation}
The solution can be written as a power series in $v,\bar{v}$ with
coefficients built out of the components of the curvature tensor
at $w_0$. Indeed, it is not hard to check that
\begin{equation}
    (\partial v(\zeta),\bar \partial \bar v(\bar \zeta ))_{\{v(\zeta),w_0\}}
    \ = \
    \sum_{n=0}^\infty \frac{(-1)^n}{2^n}
    \partial \Big(Q^n\big({\bar v(\bar \zeta)}\big) v(\zeta),
    {\bar \partial \bar v(\bar \zeta)}\Big) _{w_0},
\label{eq:metric_as_power_series_curvature_tensor}
\end{equation}
where we used the operator
\begin{equation*}
    Q(\bar{Y})X \ = \ R(X,\bar{Y})X
\end{equation*}
which is defined for arbitrary (anti-)holomorphic vectors $(\bar
Y) X $ and $R$ is the curvature tensor on our K\"ahler
supermanifold. In the case of complex projective superspace \CPn\
the curvature tensor reads
\begin{equation}
    R(X,\bar{Y})Z \ = \ (X,\bar{Y})Z+(Z,\bar{Y})X\ .
\label{eq:curvature_CP}
\end{equation}
Plugging this back in to the series
\eqref{eq:metric_as_power_series_curvature_tensor}, one may resum
the expression to obtain
\begin{equation}
 (X,\bar{Y})_{\{v,w_0\}} \ = \ \frac{(X,\bar{Y})_{w_0}}{1+ (v,\bar{v})_{w_0}} -
 \frac{(X,\bar{v})_{w_0}(v,\bar{Y})_{w_0}}{\big(1+ (v,\bar{v})_{w_0}\big)^2}
\label{eq:metric_compl_pr_space}
\end{equation}
where $X(v)$ and $\bar{Y}(\bar{v})$ are arbitrary holomorphic and,
respectively, anti-holomorphic vector fields and the scalar
product $(\phantom{c},\phantom{c})_{w_0}$ is computed with the
Fubini-Study metric~\eqref{eq:fs_metric_on_cp_blabla} at $w_0$.

In the background field method, the coordinates $v$ and $\bar v$
are now promoted to fields $v(z,\bar z)$ and $\bar v(z,\bar z)$ on
the world-sheet. The action~\eqref{eq:action_sigma} becomes
\begin{equation}
    S[v] \ = \
    \int_\Sigma d^2 z\, \left( \frac{1}{g_\sigma^2}+\frac{i\theta}{\pi}\right)
    \big(\partial v,
    \bar \partial \bar v\big)_{\{v,w_0\}} +
    \left( \frac{1}{g_\sigma^2}-\frac{i\theta}{\pi}\right)
     \big(\bar{\partial} v,
    \partial \bar{v}\big)_{\{v,w_0\}}
\label{eq:full_action_hol_coord}
\end{equation}
where the metric $(\phantom{x},\phantom{x})_{\{v,w_0\}}$ in normal
coordinates
 was computed in
eq.~\eqref{eq:metric_as_power_series_curvature_tensor} as a power
series of matrix elements of the curvature
tensor~\eqref{eq:curvature_CP}. For the applications we have in
mind, the action \eqref{eq:full_action_hol_coord} is formulated on
a world-sheet with boundary.

Let us assume that the boundary conditions that are imposed along
the boundary preserve the global supergroup symmetry. Then the
path integration factorizes into two contributions. One of them is
a finite dimensional integral along the value $w_0$ of the
fundamental field $w(z_0)$ at one point $z_0$ of the world-sheet.
The second is the path integral along its ``deviation'' $v(z)$.
For the measure, this split takes the following form
\begin{equation}
 \mathcal{D}[w,\bar{w}] \ = \  d\mu(w_0,\bar{w_0})
 \, \mathcal{D}[v,\bar{v}]\ ,
\label{eq:me_tr}
\end{equation}
where
\begin{equation}
    \mathcal{D}[v,\bar{v}]\  = \ \prod_{x\neq 0}
\frac{i^{p+q}}{2^{p+q}}\sqrt{g\big(v(x),\bar{v}(x)|w_0,\bar{w}_0\big)}
dv^1(x)\wedge d\bar{v}^1(x)\dots dv^{p+q}(x)\wedge
d\bar{v}^{p+q}(x)\ . \label{eq:xi_measure}
\end{equation}
One can check that the superdeterminant of the metric in normal
coordinates does never depend on $v(x)$. For  the Fubini-Study
metric~\eqref{eq:fs_metric_on_cp_blabla} on the complex projective
superspace \CPn\ one even finds that
\begin{equation}
g\big(v(x),\bar{v}(x)|w_0,\bar{w}_0\big) \ = \ g(w_0,\bar{w}_0)\
=\ 1\ \ . \label{eq:flat_measure}
\end{equation}
In conclusion, computations in the background field expansion for
\CPn\ are performed with the standard path integral measure using
the free field theory action
\begin{equation}
    S_0[v]\ =\
    \int_\Sigma d^2 z\, \left( \frac{1}{g_\sigma^2}+\frac{i\theta}{\pi}\right)
    \big(\partial v,
    \bar \partial \bar v\big)_{w_0} +
    \left( \frac{1}{g_\sigma^2}-\frac{i\theta}{\pi}\right)
     \big(\bar{\partial} v,
    \partial \bar{v}\big)_{w_0} \ \ .
\label{eq:free_act_with_top_term}
\end{equation}
The interaction terms are obtained by expanding the Fubini-Study
metric \eqref{eq:metric_compl_pr_space} in the fluctuation field
$v$. After this preparation we are now ready to compute some
quantities in the sigma model on \CPn.

As a warm-up example, let us calculate the index
$J^{g_\sigma,\theta}_{0,0}(q) =
Z^{g_\sigma,\theta}_{0,0}(1,1,-1;q)$, i.e.\ the partition function
of the boundary theory with $M= 0=N$ specialized to the values
$x=1=y$ and $z=-1$. It is easy to see from eq.\ \eqref{Kmod} that
the characters of Kac-modules vanish at this special point, simply
because the contributions from bosons and fermions cancel against
each other. It follows from our eq.\ \eqref{ZRinfty} that the
index $J$ vanishes at $g_\sigma = 0$. Our aim here is to show that
it actually vanishes for all values of $g_\sigma$ and $\theta$.
According to eq.~\eqref{eq:me_tr}, the perturbative partition
function $J^{g_\sigma,\theta}_{0,0}$ of the sigma model
eq.~\eqref{eq:full_action_hol_coord} can be written as
\begin{equation}
    J^{g_\sigma,\theta}_{0,0}(q)\  = \ \int
d\mu(w_0,\bar{w}_0)j^{g_\sigma,\theta}_{0,0}(w_0,\bar{w}_0).
\label{eq:complete_pf_bla}
\end{equation}
We shall call $j^{g_\sigma,\theta}_{0,0}(w_0,\bar{w}_0)$ the local
partition function. By carefully analyzing the perturbative
expansion of the partial partition function one can prove that it
receives no corrections from the interaction terms, that is
\begin{equation}
    j^{g_\sigma,\theta}_{0,0}(w_0,\bar{w}_0) \ = \
j^{(0)}_{0,0}(w_0,\bar{w}_0), \label{eq:pert_res_1}
\end{equation}
where $j^{(0)}_{0,0}(w_0,\bar{w}_0)$ is the local partition
function of the free theory~\eqref{eq:free_act_with_top_term}. The
equality~\eqref{eq:pert_res_1} may be derived with the help of the
property~\eqref{eq:metric_as_power_series_curvature_tensor} of the
metric in normal coordinates. It expresses the perturbative local
index in terms of tensor powers of the curvature tensor on \CPn.
But all the corrections to the index vanish. In fact, one may show
(see appendix \ref{sec:vanishing_invariants}) that all scalars
constructed from the tensor powers of the curvature tensor on
\CPn\ are zero. This completes the proof of eq.\
\eqref{eq:pert_res_1}. It remains to integrate the local index
over the target space coordinates $w_0$. Since neither the measure
nor the free action contain $w_0$, we infer that the local index
itself must be constant. Using that the superspace $\CPn$ has
vanishing volume we can now conclude $J^{g_\sigma,\theta}_{0,0}
(q)= 0$, as we had claimed before.

The main goal of this section is to compute 2-point functions and
thereby to determine the conformal dimensions of boundary fields
as a function of $g_\sigma$ and $\theta$. Let $\mathcal{O}[w](z)$
denote a (boundary) field of the sigma model on our K\"ahler
manifold. After insertion of the change of coordinates formula
~\eqref{eq:hol_trans_func}, the fields become functionals of the
(constant) background $w_0$ and the fluctuation field $v$. The
correlation functions are then given by
\begin{equation}
    \big\langle \prod_\nu \mathcal{O_\nu}[w](z_\nu,\bar z_\nu)
    \big\rangle \ \propto\
    \int  d\mu(w_0)   \  \Big\langle
\prod_\nu \mathcal{O_\nu}\big[c_{w_0}(v)\big](z_\nu,\bar z_\nu)
   \ e^{-S^{\mathrm{int}}_{g_\sigma,\theta}[v]}\Big\rangle_{w_0}\ .
\label{eq:def_pert_theory_bla}
\end{equation}
We compute the quantity on the the right hand side by expanding in
powers of $v$ both the interaction \emph{and} the fields
$\mathcal{O_\nu}[c_{w_0}(v)]$. The notation $\langle
\phantom{\mathcal{O}} \rangle_{w_0}$ we used in
eq.~\eqref{eq:def_pert_theory_bla} means that the expression in
brackets must be calculated in the free
theory~\eqref{eq:free_act_with_top_term} with fixed zero mode
$w_0$.

We have applied the general prescription
\eqref{eq:def_pert_theory_bla} to the computation of boundary
2-point functions for boundary condition changing fields with
$M=N$ in the \CPone\ sigma model. From the results, we obtained
the following expression for the conformal weights of tachyon
vertex operators in the representation $[k-1,0,2,0], k = 1,2,
\dots,$ of \utwo,
\begin{equation}\label{eq:Casevolg6}
h_{0,0}^{g_\sigma,\theta} [k-1,0,2,0] \ = \ \frac{g_\sigma^2}{\pi}\,
\left[1-g_\sigma^4 \left(\frac{\theta}{\pi}+2N\right)^2\right] \Cas_{\alpha = 1}[k-1,0,2,0] + O(g_\sigma^8)\ .
\end{equation}
It is easy to see \cite{Schomerus:1999ug} that conformal weights for boundary
condition changing operators with $M = N$ depend on $g_\sigma$
and $\theta$ only through the combination
\begin{equation}
    \Big(g^{eff}_\sigma\Big)^2 \ =\
\frac{g_\sigma^2}{1+g_\sigma^4\left(\frac{\theta}{\pi}+2N\right)^2}\, ,
\label{eq:eff}
\end{equation}
which gives the dependence on $g_\sigma$ and $\theta$ in the propagator of the quantum fields. 
The computation of the latter for boundary conditions of the type~\eqref{LRdef} with $M=N$ can be found in \cite{Yost}.
We have not managed to carry the computation of weights to higher
orders. This is partly due to the fact that the background field
expansion breaks the psl(2$|$2) symmetry down to sl(1$|$2) so that
some of the simplifications that arise from special features of
the Lie superalgebra psl(2$|$2) (see e.g.\
\cite{Bershadsky:1999hk}) are not directly applicable.
Nevertheless, we take eq.\ \eqref{eq:Casevolg6} as a strong
indication that boundary weights of tachyonic vertex operators
transforming in some representation $\Lambda$ of \utwo\ behave as,
\begin{equation}\label{last_eq_label}
h_{M,N}^{g_\sigma,\theta} (\Lambda) \ = \
h^\ast_{M,N}(g_\sigma,\theta) + \frac{g_{M,N}(g_\sigma,\theta)}{4}
\Cas_{\alpha=1}(\Lambda)
\end{equation}
with some functions $h^\ast_{M,N}(g_\sigma,\theta)$ and
$g_{M,N}(g_\sigma,\theta)$ that will be determined below. This
conjectured behavior of the conformal weights will be one of the
central ingredients in our formula for the boundary partition
function of the \CPone\ model. It has also passed extensive
numerical checks that we describe in the second part of this work.

\subsection{Partition function at finite coupling}

It is now time to spell out the central formula of this paper. We
propose the following boundary partition function of the \CPone\
model with monopole bundle boundary conditions $M,N$ imposed along
the two boundaries of the strip,
\begin{eqnarray}
Z^{g_\sigma,\theta}_{M,N}(x,y,z;q) & = & q^{\frac12
\lambda_{M,N}(g_\sigma,\theta) (\lambda_{M,N}(g_\sigma,\theta)-1)}
\sum \chi^K_{[j_1,j_2,a,l/2]}(x,y,z) \, \times \\[4mm]
& & \hspace*{.5cm} \times \ q^{\frac{1}{4}
g_{M,N}(g_\sigma,\theta)\, \delta_l C^{(2)}([j_1,j_2,a,l/2])}\
\psi^{K}_{\left[j_1,j_2,a,l/2\right]}(q) \ \ . \nonumber
\label{ZRfinite}
\end{eqnarray}
The partition function depends on the couplings $g_\sigma$ and $\theta$ through the functions
$\lambda_{M, N}(g_\sigma, \theta)$ and
$g_{M,N}(g_\sigma, \theta)$. These functions are universal, i.e.\
do not depend on the representation $[j_1,j_2,a,b]$ the field
transforms in. We will provide explicit formulas below (see eqs.\
\eqref{lform} and \eqref{flrel}). The functions $\lambda$ and $g$
also turn out to be the same for all \CPn\ models, regardless of
the value of $S$. Hence, our partition function depends on $S$
only through the branching functions $\psi$ and a certain
difference $\delta_l C^{(2)}$ of Casimir elements of u(S$|$S). For
$S=2$, the former were determined in section 3 through our
analysis of the model at $g_\sigma =0$. The relevant Casimir
element $\Cas_{\alpha}$ was displayed in eq.\
\eqref{eq:Casimiralpha} before. What appears in eq.\
\eqref{ZRfinite} is the difference
\begin{equation} \label{eq:deltaC}
\begin{split}
  \delta_l C^{(2)}([j_1,j_2,a,l/2]) & = \
     \Cas_{\alpha=1}([j_1,j_2,a,l/2]) - \Cas_{\alpha=1}(\Lambda_{0,l})
    \\[2mm]  & = \ 2[j_1(j_1+1)- j_2(j_2+1)] + l(a-2)
    - l^2\ + |l|
\end{split}
\end{equation}
The weight $\Lambda_{0,l}$ corresponds to the representation of
the ground state. The latter minimizes the value of
$-\Cas_{\alpha=1}$ among all the representations that appear in
the decomposition \eqref{ZRinfty}, see appendix
\ref{sec:Laplacian} for details.

Let us now address the two functions $\lambda$ and $g$ in more
detail. Obviously, the function $\lambda$ determines the conformal
weight of the ground state in the boundary theory. The function
$g$, on the other hand, encodes how conformal weights of the
excited states change relative to the ground state as we vary the
two bulk couplings $g_\sigma$ and $\theta$. We claim that both
$\lambda$ and $g$ are independent of the integer $S$, i.e.\ they
are the same for all projective superspaces \CPn. We shall only
sketch the argument here. It is based on the observation that all
\CPn\ models contain symplectic fermions as a true subsector \cite{SaleurReadi}. In
other words, all fields of the symplectic fermion model \CPzero\
can be embedded into the theory with target space \CPn\ in such a
way that their correlation functions are preserved under the
embedding. A very elegant proof of this statement will be given in
a forthcoming publication. For the \CPone\ model, states from the
symplectic fermion subsector are to be found within the first two
(atypical) multiplets in the decomposition \eqref{Z0part} of
fields with weight $h=0$ at $g_\sigma=0$. Since the weights of
theses two multiplets determine the two functions $\lambda$ and
$g$ uniquely, we can compute both $\lambda$ and $g$ within the
free field theory of symplectic fermions.

Our first goal now is to compute the functions $\lambda_{M,N}$
within the symplectic fermion model. To this end we look back at
our formula \eqref{glue} that describes the gluing condition of
fields at the boundary in terms of the parameters $N,M$ and
$\theta$. These boundary conditions are of Neumann type, twisted
by the presence of a nontrivial matrix $W$ of the form
$$
W(\Theta) = i g_\sigma^2 \left( \begin{matrix}  \Theta& 0
\\ 0 & -\Theta   \end{matrix}\right)\ .   $$
The matrix $W$ relates the derivatives along and perpendicular to
the boundary of the world-sheet. Since $\Theta = 2N + \theta/\pi$,
the matrix $W$ may be written as a sum $W = B(\theta) + F(N)$ of a
`bulk magnetic field' $B = B(\theta)$ and the `field strength' $F
= F(N)$ of the monopole. If we choose different monopole numbers
$M,N$ on the two sides of the strip, the gluing conditions along
the left and the right boundary are different. Consequently, the
corresponding boundary condition changing fields must be in
twisted sectors. In order to determine the twist parameter
$\lambda$, we reformulate the boundary condition in terms of a
gluing automorphism $\Omega$ that relates chiral fields rather
than the derivatives $\partial_x$ and $\partial_y$. The gluing
automorphism is given by
$$ \Omega = \frac{1+W}{1-W} \ \ . $$
Let us denote the two different values of $\Theta$ along the left
and the right boundary by $\Theta_1$ and $\Theta_2$. Similarly, we
shall use the symbols $W_i = W(\Theta_i) $ and $\Omega_i =
\Omega(\Theta_i)$ for the corresponding field strength $W$ and the
gluing automorphism $\Omega$ along the two half-lines. It follows
that the symplectic fermions possess monodromy
$$ \Omega_{12} = \Omega_1 \Omega_2^{-1} \ =\
\frac{\kappa +W(\Theta_1-\Theta_2)}{\kappa - W(\Theta_1-\Theta_2)}
 \ \ \ \mbox{where} \ \ \
 \kappa  \ = \ 1 + g_\sigma^4\Theta_1\Theta_2\
$$
when taken around a boundary field insertion. The trace of this
monodromy matrix $\Omega_{12}$ determines the twist parameter of
the symplectic fermions through $2\cos 2\pi \lambda = \tr
\Omega_{12}$. Putting all this together we find
\begin{equation}
\label{lform}
 \cos 2\pi
\lambda_{M,N}(g_\sigma,\theta) \ = \ \frac{(1 + g_\sigma^4
\Theta_1\Theta_2)^2 - (\Theta_{1}-\Theta_2)^2 g_\sigma^4}{ (1 +
g_\sigma^4 \Theta_1\Theta_2)^2 + (\Theta_{1}-\Theta_2)^2
g_\sigma^4}\ ,
\end{equation}
where $\Theta_1 = 2M + \theta/\pi$ and $\Theta_2 = 2N +
\theta/\pi$. There are a few special cases to be discussed. To
begin with let us choose $M=N$. When the two boundary conditions
on both sides of the interval are identical so that $\Theta_1 =
\Theta_2$, then $\cos 2\pi \lambda = 1$ and consequently  the
twist parameter vanishes. Similarly, we note that the twist
parameter always vanishes in the limit of infinite radius, i.e.\
when $g_\sigma = 0$. The boundary theory with vanishing twist
parameter was constructed explicitly in \cite{Creutzig:2006wk}.
The more general case has been considered in
\cite{Creutzig:2008an}.

It remains to find the second set of functions $g_{M,N}$. We shall
see momentarily that they are very closely related to
$\lambda_{M,N}$. As we have just argued, the ground states in our
symplectic fermion model on the upper half-plane are twist fields
with a twist parameter $\lambda$. The corresponding conformal
weight is
$$ h^{\text{gr}}_\lambda \ = \ \frac12\lambda (\lambda-1)\ \ . $$
Excited states in the symplectic fermion model are generated by
acting with modes of the form $\chi_{-\lambda-n}, n \leq 0$.
Hence, the first excitations above the ground states possess
conformal weight $h^{\text{ex}} = h_\lambda^{\text{gr}} +
\lambda$. These states of the symplectic fermions are embedded
into the second term in the decomposition \eqref{Z0part}.
Consequently, the two functions $\lambda$ and $g$ must be related
by
\begin{equation} \label{flrel}
\lambda_{M,N}(g_\sigma,\theta) \ = \ \frac{1}{4}\delta
C^{(2)}(\Lambda_{1,M-N}) g_{M,N}(g_\sigma,\theta) \ = \
\frac{1}{2} |M-N| g _{M,N}(g_\sigma,\theta)\
\end{equation}
where the weights $\Lambda_{0,l}$ and $\Lambda_{1,l}$ in terms of
the labels $[j_1,j_2,a,b]$ can be found in
sec.~\ref{sec:Laplacian}. The equation determines $g_{M,N}$ in
terms of the twist parameter $\lambda_{M,N}$, at least when $M
\neq N$. When $M=N$, the twist parameter vanishes. Since the
coefficient $|M-N|$ on the right hand side of equation
\eqref{flrel} also goes to zero as $M \rightarrow N$, the function
$g_{N,N}$ can be computed as
\begin{equation} \label{eq:gNN}
g_{N,N}(g_\sigma,\theta) \ = \ \lim_{M\rightarrow N} \left(
\frac{2\lambda_{M,N}(g_\sigma,\theta)}{|M-N|} \right) \ = \
\frac{4g_\sigma^2}{\pi\big[1+ g_\sigma^4
(\frac{\theta}{\pi}+2N)^2\big]}\ .
\end{equation}
Hence, the universal function $g_{N,N}$ is related to
the effective coupling $g^{eff}_\sigma$ we found while analyzing
the background field expansion in eq.\ \eqref{eq:eff},
\begin{equation} \label{eq:g00}
g_{N,N}(g_\sigma,\theta) \ = \ \frac{4}{\pi} \,
\left(g_\sigma^{eff}\right)^2\ \ . \end{equation}
%

Before we conclude this subsection let us spell out one more
special case of our expression for $\lambda$ to prepare for our
lattice analysis in the next section. In the second part, we will
perform numerical calculations for nonzero values of the monopole
charges $M,N$. Simulations with $M=0$ and $N=-1$ at the point
$g_\sigma = 1$ will give the ground state energy $h_\lambda =
-1/8$. This corresponds to the twist parameter $\lambda = 1/2$. To
reproduce this values, we need
$$ \cos 2 \pi \lambda_{0,1}(g_\sigma^2 =1,\theta) \ =\
\frac{\left(1+ \frac{\theta}{\pi}\left(
\frac{\theta}{\pi}-2\right)\right)^2 - 4}{\left(1+
\frac{\theta}{\pi}\left(\frac{\theta}{\pi}-2\right)\right)^2 +
4} \ = \ -1\ \ .
$$
We read off that the lattice model must flow to the continuum
theory with $\theta = \pi$. It is interesting to note that the
$\theta$ angle of the bulk theory may be determined from the
behavior of boundary conformal weights.
\newpage
\noindent
{\huge{\bf Part II: Discretization and Numerics} }\\[2cm]
Our proposal for the exact partition function of \utwo\ symmetric
boundary theories is based on two central ingredients. On the one
hand, there are perturbative studies around $g_\sigma =0$ that
indicate that conformal weights evolve with the quadratic Casimir
element. In addition, the close relation of the \CPone\ model with
symplectic fermions allowed us to determine the universal
functions $g_{M,N}$ in front of the Casimir element and the ground
state energies. While the embedding of symplectic fermions is a
non-perturbative feature of the \CPone\ model, the Casimir
evolution was only analyzed perturbatively in the coupling
constant $g_\sigma$. In order to further test our formulas for the
evolution of conformal weights, we shall now introduce a lattice
model. The discrete theory can be studied numerically without any
need to expand in the coupling constant $g_\sigma$. We shall find
remarkable agreement between our analytical studies of the
continuum model and the numerical results for its discretization.
The agreement suggests that our proposal for the partition
functions of boundary theories is exact. In particular, it does
not seem to receive non-perturbative corrections.

\section{Brauer algebra and alternating \u\ spin chain}

The main purpose of this section is to establish the Hamiltonian
\eqref{eq:general_ham} with $a=0$ as a promising candidate to
describe a discrete version of the bulk dynamics in the \CPn\
models. Our discussion will require some background on (walled)
Brauer algebras which we describe first.

Lattice studies of two dimensional   \bCPn\ models  involve, in
their most direct version, the Monte Carlo study of a model with
$S$ dimensional complex unit vectors on the vertices and
$\ensuremath{\text{U}(1)}$ gauge fields on the edges of a square
lattice, together with the proper discretization of the
topological term (this is somewhat less obvious of course, as
there is no topology on the lattice) \cite{BergLuscher,Seiberg}.
Condensed matter physics has provided an alternative to this
approach, where the fields can now emerge dynamically as
collective excitations of quantum spins. The conjecture by Haldane
\cite{Haldane,Affleck} that the long distance properties of
$\ensuremath{\text{SU}(2)}$  spin chains is described by the
$\ensuremath{\text{O}(3)}$ sigma model at $\theta=0$ for integer
($\theta=\pi$ for half integer) spin opened the way to studying
the mapping of most general spin chains to sigma models
\cite{ReadSachdev}. Lately, this idea has been intensely revisited
in the context of the AdS/CFT duality.

Geometric quantization arguments \cite{ReadSachdev,Wiegmann} show
that the simplest spin chain we could use to understand the \CPn\
model is based on alternating  the fundamental  representation $V$
of \u\ and the dual fundamental $V^\star$.
For a description of these modules, see \cite{Read:2001pz}.
Moreover, for a homogeneous chain, we should get $\theta=\pi$.

Integrable spin chains for this choice of representations turn out
to have a non generic continuum limit, described  by  a WZW model
\cite{SaleurSchomerus}. To see the physics of the \CPn\ model, we
need to use more generic interactions. The ones we shall find to
describe the physics of the continuum theory do not preserve
integrability. Fortunately, a lot can still be understood
analytically by studying the properties of the chains under the
simultaneous action of the (super) Lie algebra symmetry and its
commutant \cite{SaleurRead}. In the present case, this commutant
is given by the walled Brauer algebra. The algebraic approach that
we are about to review has a number of appealing features. In
particular, up to a certain point of the analysis, it may be
formulated without any reference to the value of $S$.

\begin{figure}[t]%
\centerline{\includegraphics[scale=0.8]{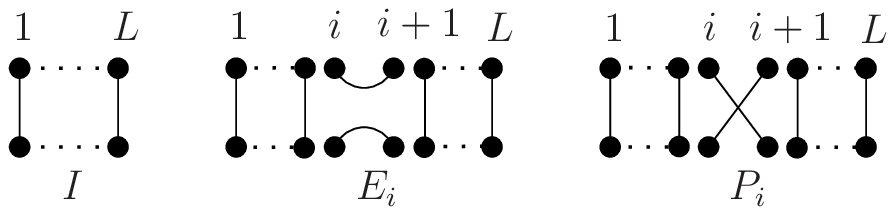}
\hspace{2cm}
\includegraphics[scale=0.8]{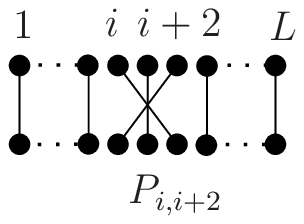}}%
\caption{The identity $I$ and the generators $E_i,P_i$ of the Brauer algebra of
dimension
$(2L-1)!!$ are represented on the left; the walled Brauer algebra generator
$P_{i,i+2} = P_i P_{i+1}P_i$
is represented on the right.}%
\label{fig:gen_br_alg}%
\end{figure}

Throughout the following subsections we denote the generators of
the Brauer algebra $B_{2L}(0)$ by $E_1,P_1,\dots,E_{2L-1},P_{2L-1}$.
In the symbol $B_{2L}(0)$, the index $2L$ is related to the
dimension of the Brauer algebra by $\dim B_{2L}(0) = (4L-1)!!$ and the
parameter in parenthesis denotes the so-called fugacity for loops.
The defining relations of $B_{2L}(0)$ can be found in \cite{ram}.
The words of this Brauer algebra admit a representation as graphs
on $4L$ labelled vertices with $2L$ edges connecting the vertices
pairwise in all $(4L-1)!!$ possible ways (crossings are allowed).
The identity $I$ of the Brauer algebra and the generators $E_i,P_i$
are represented by the graphs on the left in
fig.~\ref{fig:gen_br_alg}.
In order to multiply the diagrams one arranges the first $2L$
vertices horizontally with the remaining $2L$ vertices on top of
the first ones. The product of a diagram $d_1$ with a diagram
$d_2$ is the diagram $d_1 d_2$ obtained by i) placing the diagram
$d_1$ on top of the diagram $d_2$, ii) identifying the top of the
diagram $d_2$ with the bottom of the diagram $d_1$ and iii)
replacing every loop generated in this process by 0. The periodic
Brauer algebra is an extension of the Brauer algebra by two
generators $E_{2L}$ and $P_{2L}$ which satisfy the same defining
relation as the generators of the Brauer algebra if the index
$i\equiv i+2L$ is regarded as periodic. The words of the periodic
Brauer algebra are diagrams with the top and the bottom being
circles wrapped around a cylinder and carrying $2L$ vertices each,
such that the latter are pairwise connected in all possible ways
by $2L$ edges  living on the surface of the cylinder. The periodic
Brauer algebra has infinite dimension.

The elements $E_i$ and $P_{i,i+2}= P_i P_{i+1}P_i$ freely generate
a subalgebra called the walled Brauer algebra. The generators
$P_{i,i+2}$ are represented on the right in
fig.~\ref{fig:gen_br_alg}. This walled Brauer algebra is of
central importance for the study of $\gl(S|S)$-invariants as
explained in the following. Let $V$ denote the fundamental
representation of $\gl(S|S)$ and $V^\star$ be its dual.
Generalizing the well known statement for $\gl(N)$, Sergeev proved
\cite{Sergeev} that there is a surjective homomorphism from the
walled Brauer algebra to the invariants of the tensor module
$(V\otimes~ V^\star)^{\otimes 2L}$ or, equivalently, to the
$\gl(S|S)$-centralizer of  $(V\otimes~ V^\star)^{\otimes L}$. This
means that the $\gl(S|S)$-centralizer of $(V\otimes~
V^\star)^{\otimes L}$ can be viewed as some representation of the
walled Brauer algebra. In particular, the most general
$\gl(S|S)$-symmetric spin chain Hamiltonian $H$ one can write down
must represent some element of the walled Brauer algebra. If we
restrict to nearest neighbor interactions only (hence defining a
$\gl(S|S)$ version of the Heisenberg chain),  we get a Hamiltonian
of the form
\begin{equation}
    H_{\mathrm{TL}}\ =\  - \sum_i t_i E_i\ .
\label{eq:gen_TL_ham}
\end{equation}
This Hamiltonian, all of its powers and the corresponding
evolution operator $e^{-\tau H_{\mathrm{TL}}}$ lie entirely in the
Temperley-Lieb subalgebra of the walled Brauer algebra. Thus, by
the double centralizer theorem, the  symmetry of $H_{\mathrm{TL}}$
must be bigger then $\gl(S|S)$. One can show \cite{saleur_young}
that the spectrum of low lying excitation of the homogeneous chain
$H_{\mathrm{TL}}$ in the scaling limit is described by the free
field theory of a pair of free symplectic fermions,
\begin{equation}
S \ \sim \ \int d^2z\, \partial_\mu \eta_1(z,\bar z
)\partial^\mu\eta_2(z,\bar z). \label{eq:free_SF_action}
\end{equation}
The degeneracies of the excitations of the lattice model must be
computed by employing independent representation theoretic tools
developed in \cite{SaleurRead}.

We are naturally interested in deformations of the Temperley-Lieb
Hamiltonian~\eqref{eq:gen_TL_ham} which break the symmetry all the
way down to $\gl(S|S)$ and preserve conformal invariance in the
continuum limit. The simplest $\gl(S|S)$-symmetric Hamiltonian is
the sum of generators of the walled Brauer algebra. Since the
generator $P_{i,i+2}$ corresponds to second nearest neighbor
interactions on the spin chain $(V\otimes~ V^\star)^{\otimes L}$,
it is natural to consider the following $\gl(S|S)$-symmetric
deformation of the Hamiltonian \eqref{eq:gen_TL_ham}
\begin{equation}
  H_{\text{gen}} \ = \   -\sum_i \,\big[ t_i E_i + w_i P_{i,i+2} +
  a_i E_i E_{i+1} + b_i E_{i+1}E_i\big]\ .
\label{eq:gen_WB_ham}
\end{equation}
The eigenvalues of the Hamiltonian~\eqref{eq:gen_WB_ham} are more
easily computed by working in the adjoint -  that is in the
diagrammatic - representation of the walled Brauer algebra, rather
than in the representation on $(V\otimes~ V^\star)^{\otimes L}$.
However, when switching between the alternating spin chain and
adjoint representations of the walled Brauer algebra one looses
control of the degeneracies of eigenvalues. These can be recovered
from representation theory by methods similar to those used in
\cite{Candu:2008vw}. We shall call the
Hamiltonian~\eqref{eq:gen_WB_ham} algebraic when it is considered
in the adjoint representation of the walled Brauer algebra. The
actual spectrum of the $\gl(S|S)$ alternating spin chain will be a
subset of the spectrum of the algebraic
Hamiltonian~\eqref{eq:gen_WB_ham}. We call this subset a
$\gl(S|S)$-sector of the algebraic Hamiltonian. With a little bit
of representation theory of the walled Brauer algebra one can
prove that the eigenvalues of the $\gl(S-1|S-1)$ spin chain
Hamiltonian are a subset of the eigenvalues of the $\gl(S|S)$ spin
chain Hamiltonian. This is done in essentially the same way as in
\cite{Candu:2008vw}.

At a critical point, the space of states of the statistical model
usually possesses some additional discrete symmetries. Without loss
of generality one can impose these discrete symmetries on the
Hamiltonian~\eqref{eq:gen_WB_ham}, thereby reducing the number of
independent couplings $t_i,w_i,a_i,b_i$. The scale invariant
vacuum in periodic boundary conditions is necessarily translation
invariant.  Consequently, we shall restrict to homogeneous
Hamiltonians~\eqref{eq:gen_WB_ham}, i.e.\ to Hamiltonians that are
invariant under the discrete shift automorphism
\begin{equation*}
    E_i \rightarrow E_{i+1}, \quad P_{i-1,i+1}\rightarrow P_{i,i+2}
\end{equation*}
of the periodic walled Brauer algebra. If we additionally assume
invariance with respect to the reflection automorphism
\begin{equation*}
    E_i \rightarrow E_{2L-i+1},\qquad P_{i,i+2}\rightarrow P_{2L-i,2L-i+2},
\end{equation*}
then the Hamiltonian becomes
\begin{equation}
    H \ =\  - \sum_{i=1}^{2L}\, \big[ t E_i + w P_{i,i+2} + a (E_i E_{i+1}+
    E_{i+1}E_i)\big]\ .
\label{eq:general_ham}
\end{equation}
We shall restrict to real couplings $t,w$ and $a$. It will take
some more discussion to gain sufficient intuition into the new
couplings $w$ and $a$. In particular we shall argue that $w$ is an
exactly marginal coupling which corresponds to the radius
parameter $R$ of the continuum theory. The coupling $a$, on the
other hand, seems to have little effect and will eventually be set
to zero.

In order to interpret the couplings $a$ and $w$ we shall mostly
work with the $\gl (1|1)$ subsector, i.e.\ we will consider the
Hamiltonian \eqref{eq:general_ham} as an operator on the state
space of the $\gl(1|1)$ alternating spin chain. The resulting
theory is a discrete version of the free theory of symplectic
fermions. We can make the link by introducing a set of $2L$
creation and annihilation fermionic operators
\begin{equation}
    \{\varphi_i,\bar{\varphi}_j\} \ = \ \delta_{ij},\qquad
    i,j=1,\dots, 2L\ .
\label{eq:discrete_fermions}
\end{equation}
These may be employed to represent the generators of the walled
Brauer algebra through the following quadratic expressions
\begin{align}\label{eq:wba11}
    E_j &= \ (-1)^j ( \bar{\varphi}_j-\bar{\varphi}_{j+1} ) (
    \varphi_j+\varphi_{j+1} ) \\[2mm]
    P_{j-1,j+1} &= \ (-1)^j\big[1- (
    \bar{\varphi}_{j-1}-\bar{\varphi}_{j+1} ) (
    \varphi_{j-1}-\varphi_{j+1} )\big]\ .\notag
\end{align}
The continuum limit of the $\gl(1|1)$ Hamiltonian
\eqref{eq:general_ham} with $a=0$ is described by an action of the
type~\eqref{eq:free_SF_action}, the same we found for $w=0$. In
other words, when $a=0$ and $S=1$, the perturbation with
$P_{i,i+2}$ is truly redundant: On the lattice, its only effect is
to renormalize the sound velocity
\begin{equation*}
    v_{\text{sound}} \ = \ 2t\sqrt{1+4w} \ .
\end{equation*}
Switching on the coupling $a\neq 0$ in the $\gl(1|1)$ alternating
spin chain~\eqref{eq:general_ham} provides a quartic interaction
in terms of the discrete fermions~\eqref{eq:discrete_fermions}.
The resulting model does not seem to be exactly solvable. One of
the fourth order terms of the continuum theory,
\begin{equation*}
 \delta S \sim  \int d^2z\,
\eta_1(z)\eta_2(z)\partial_\mu\eta_1(z)\partial_\mu\eta_2(z),
\end{equation*}
has been studied in detail in~\cite{arboreal_gaz}. It was shown to
be either marginally relevant or marginally irrelevant, depending
on the sign of its coupling. In the continuum theory, adding a
fourth order term in the fermions is actually inconsistent with the
$\gl (1|1)$ symmetry of the model.\footnote{We thank N. Read for a
discussion of this point.} Free symplectic fermions possess 16
bulk fields of weight $h=\bar{h}=1$. These are obtained by
multiplying $1,\eta_1,\eta_2,\eta_1\eta_2$ with $\partial\eta_1$
or $\partial\eta_2$ and a similar term with $ \bar\partial$ in
place of $\partial$. Under the right (or left) action of
$\gl(1|1)$, these transform in four indecomposable projectives. A
closer look reveals that only two of the 16 fields are true
invariants, i.e.\ they are annihilated by all the $\gl (1|1)$
generators. These two fields are quadratic in the fermions. Hence,
adding a fourth order term to the symplectic fermion model breaks
the $\gl(1|1)$ symmetry. We thus conclude that non-zero values of
the parameter $a$ in the lattice theory will not effect the
continuum theory, at least not for small enough value of $a$.

We suggest that the above conclusions should essentially remain
correct for $S > 1$. Numerical diagonalization of the algebraic
Hamiltonian~\eqref{eq:general_ham} for $a=0$ indicates that its
lowest eigenvalue lies in the $\gl(1|1)$-sector. This means that
one can compute this lowest eigenvalue by restricting the
algebraic Hamiltonian~\eqref{eq:general_ham} to the state space of
the $\gl(1|1)$ alternating spin chain. Hence, $w$ should be
exactly  marginal even for $S > 1$, at least as long as $a = 0$.
It is tempting to think that this conclusion remains valid for
nonzero values of $a$ and that $a$ continues to be irrelevant.
 
To have a complete correspondence between the couplings of the $\CPn$ sigma model and those of our lattice model we are still left with the problem to identify a second lattice coupling that could implement the $\theta$ angle. Let us
anticipate that the $\theta$ parameter corresponds to staggering
the couplings of the lattice model. We will get back to this in
the conclusion.

In the following we shall provide strong evidence for our claim
that the spectrum of low lying excitations of the alternating
$\gl(S|S)$ spin chain~\eqref{eq:general_ham} with $a=0$ is
described by the sigma model on the complex projective superspace
$\CPn$ with $\theta=\pi$.

\section{Open alternating $\gl(S|S)$ spin chain}\label{sec:untwisted}

Following the outcome of our discussion in the last section, let
us now work with the alternating $\gl(S|S)$-spin chain on the
space $(V\otimes~ V^\star)^{\otimes L}$ with Hamiltonian
\begin{equation}
    H \ =\  -\sum_{i=1}^{2L-1}E_i - w \sum_{i=1}^{2L-2}P_{i,i+2}\ .
\label{eq:open_untwisted_chain_ham}
\end{equation}
In order to compare numerical results with the continuum theory,
we need to consider an open chain. Numerical evidence suggests
that in the limit $w\to \infty$ and $L\to \infty$, the eigenvalues
$E_h(L)$ of the Hamiltonian~\eqref{eq:open_untwisted_chain_ham}
become infinitely degenerated. Therefore, we identify the point $w
= \infty$ with the large volume limit of the sigma model on the
complex projective superspace $\CPn$. A similar identification has
been proposed in \cite{Candu:2008vw} for the $\OSp(2S+2|2S)$-spin chain
on $V^{\otimes L}$.

Without any additional algebraic guidance, the spectrum of the
Hamiltonian~\eqref{eq:open_untwisted_chain_ham} is rather
difficult to analyze. In order to unravel some of the structure,
it is useful to classify eigenvalues according to the
representations of the walled Brauer algebra that appear in the
decomposition of the state space $(V\otimes~ V^\star)^{\otimes
L}$. If one is interested in states that transform according to
some irreducible representation of the $\gl(S|S)$ symmetry, it
pays off to identify those representations of the walled Brauer
algebra that are compatible with the required symmetry. The
Hamiltonian~\eqref{eq:open_untwisted_chain_ham} may then be
restricted and diagonalized within each such building block.

We shall be mainly concerned with the numerical analysis of
excitations of $H$ whose eigenvalues vanish in the limit
$w\to\infty$. On the sigma model side, these are the scaling
dimensions of tachyonic fields, i.e.\ of those fields that can be
built from square integrable functions on the complex projective
superspace $\CPn$. According to the results of \cite{Zhang} (see
also part I of this work), the space of tachyonic fields may be
identified with the multiplicity free direct sum of
supersymmetric, self-dual, traceless $\gl(S|S)$ tensors $t(k,k)$
of rank $2k>2$ and the indecomposable traceless but reducible
tensor $t(1,1)=V\otimes~ V^\star$. Note that the trivial tensor
$t(0,0)$ is a submodule of $V\otimes~ V^\star$. In our analysis of
the \CPone\ model, these were denoted by $t(k,k) = \Lambda_{k,0}$
for $k \geq 2$. The space $t(1,1)$ contains $\Lambda_{1,0}$ and
the trivial module $\Lambda_{0,0}$ twice. More details on these labels
can be found in appendix B.

We now restrict the
Hamiltonian~\eqref{eq:open_untwisted_chain_ham} to the submodule
of $(V\otimes~ V^\star)^{\otimes L}$ that contains all states in
the $\gl(S|S)$ representations $t(k,k)$, where $k=0,\dots, L$. The
vector space of all possible embeddings of $\gl(S|S)$ tensors
$t(k,k)$ into $(V\otimes~ V^\star)^{\otimes L}, k\neq 1,$ can be
endowed with an action of the walled Brauer algebra and it
provides an (indecomposable) representation which we denote by
$T_{L,L}(k,k)$. Similarly, the vector space of all possible
embeddings of the $\gl(S|S)$ tensor $t(1,1) = V\otimes~ V^\star$
into $(V\otimes~ V^\star)^{\otimes L}$ can be endowed with an
action of the walled Brauer algebra. In this case, the space gives
rise to an indecomposable representation $I_{L,L}$. It is not
difficult to see that the space $T_{L,L}(0,0)$ (which we defined
previously) is a submodule of $I_{L,L}$. The corresponding
quotient will be denoted by $T_{L,L}(1,1)=I_{L,L}/T_{L,L}(0,0)$.
The space $T_{L,L}(1,1)$ is actually not irreducible either. In
fact, it can be shown to possess the module $T_{L,L}(0,0)$ as a
quotient. All these statements may be proved using the geometric
(adjoint) representation of the walled Brauer algebra.

\begin{figure}
\centerline{\includegraphics[scale=0.7]{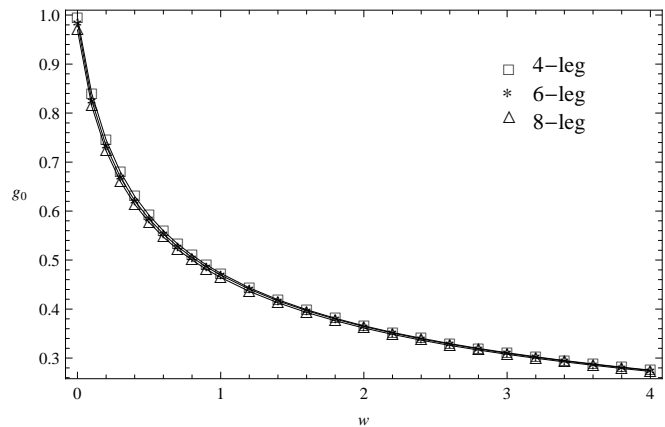}}%
\caption{Plot of $g_{0,0}(w)$ extracted from the watermelon
exponents $h_{0,0}(2)$, $h_{0,0}(3)$ and $h_{0,0}(4)$
computed at $L=9$ with the help of eq.~\eqref{eq:cas_ev_untwisted}.}%
\label{fig:g_even_open}
\end{figure}

Borrowing from the literature on self-avoiding walks,  we shall
call the lowest eigenvalue of the
Hamiltonian~\eqref{eq:open_untwisted_chain_ham} in the space
$T_{L,L}(k,k)$ the {\em $2k$-legged water melon exponent} $h_{0,0}
(k)$. According to our discussion above, the degeneracy of
$h_{0,0}(k)$ is $\dim t(k,k)$. The numerical results presented in
fig.~\ref{fig:g_even_open} strongly suggest that the continuum
limit of the  watermelon exponents is given by the very simple
expression
\begin{equation}
    h_{0,0}(k) \ = \ \frac{g_{0,0}(w) \Cas(k)}{4} \ = \
    \frac{g_{0,0}(w) k(k-1)}{2}\ ,
\label{eq:cas_ev_untwisted}
\end{equation}
where $\Cas(k)$ is the value of the quadratic Casimir\footnote{ For these
representations, the value of the quadratic Casimir is independent of
$\alpha$, see (\ref{eq:cas_eg_standard}).} in the
irreducible representations $t(k,k)$ for $k\neq 1$. For $k=1$,
$\Cas(k) = 0$ is the value of the quadratic Casimir in either the
adjoint or the trivial representation of $\gl(S|S)$. The
degeneracy of the $h_{0,0}(1)$ watermelon exponent with the vacuum
is due to the fact that, as we mentioned above, $T_{L,L}(0,0)$ is
a quotient of $T_{L,L}(1,1)$.

The numerical results should be compared with our formulas
\eqref{eq:deltaC} and \eqref{eq:g00} that determine the conformal
weight $h = \delta C^{(2)} g_{0,0}/4$  of boundary fields in the
continuum model. Using the association of the $k^{\text{th}}$ watermelon
exponent with the weight $\Lambda_{k,0}$ and the dictionary at the
end of appendix B, we conclude that
$$ \delta_0 C^{(2)}([k-1,0,2,0]) \ = \ 2k(k-1) \ \ . $$
This is in perfect agreement with our continuum theory. Note that
both on the lattice and in the continuum the ratio between the
conformal weight and the value of the Casimir element is
universal, i.e.\ it is independent of $k$. On the lattice, the
universal function $g_{0,0} = g_{0,0}(w)$ depends on the lattice
coupling $w$. The corresponding function $g_{0,0}=
g_{0,0}(g_\sigma,\theta)$ is known explicitly, see eq.\
\eqref{eq:g00}. Anticipating that $\theta =\pi$ in the continuum
limit of our lattice theory (see below), we can use the
identification $g_{0,0}(w) = g_{0,0}(g_\sigma,\theta=\pi)$ to
determine the functional dependence $w = w(g_\sigma)$ of the
lattice on the sigma model coupling $g_\sigma$.

\section{Twisted open alternating $\gl(S|S)$ spin chains}

The numerical analysis performed in the previous section suggests
that the spectrum of the open $\gl(S|S)$ spin chain is described
in the continuum limit by the sigma model on $\CPn$ subject to
Neumann boundary conditions or modified Neumann boundary
conditions in the presence of a $\theta$-term. However, the sigma
model on $\CPn$ admits a much larger set of boundary conditions
that do not break the global $\gl(S|S)$ symmetry, namely those
described by the nontrivial complex line bundles over $\CPn$. The
complex line bundles can be different at the different ends of the
string and we label them by two integers $M$ and $N$ called
monopole charges. These bundles may be introduced by adding
boundary terms to the action, that is integrals of locally defined
1-forms along the two boundaries. Each of these forms is then
interpreted as a connection defining a complex line bundle.
Naturally, if the two bundles are different, then so are the
boundary conditions at the two boundaries of the world-sheet.
Twisting of the spectrum should then be expected when $M\neq N$.
In fact, as we showed in sec.~4.2, the $\gl(1|1)$ subsector of the
$\CPn$ sigma model is described by a pair of twisted free
symplectic fermions with twisting parameter
\begin{equation}\label{eq:sm_tw_pred}
  \tan \pi \lambda_{M,N} = \frac{2lg_\sigma^2}
  {1 + \Theta_1\Theta_2g_\sigma^4},
\end{equation}
where $\Theta_1 = \tfrac{\theta}{\pi}+2M$, $\Theta_2
=\tfrac{\theta}{\pi}+2N$ and $l=M-N$. It is natural to ask if one
can associate a spin chain to each of these more general boundary
conditions. As we explain in the following, this is indeed
possible. We shall describe the general setup in the following
subsection. Then we describe our numerical results, first for the
$\gl(1|1)$ subsector and then for the watermelon exponents in the
general twisted open chain.

\subsection{Monopole boundary conditions}
\label{sec:discrete_monopole}

The space of sections in the non-trivial complex line bundles over
$\CPn$ is endowed with an action of $\gl(S|S)$ rather than
$\psl(S|S)$. Therefore, in order to break the $\psl(S|S)$ symmetry
one can proceed by considering the chain of
sec.~\ref{sec:untwisted} with some extra $V$'s or some extra
$V^\star$ attached to the ends of the chain. Depending on what we
attach to either end of the chain, there are four cases to
consider. We list them in the following together with the
Hamiltonians we chose to describe their dynamics
\begin{equation}\label{eq:tw_chain_general}
\begin{array}{rlll}
 V^{\otimes m} & \!\!\otimes\,   (V\otimes V^\star)^{\otimes L}
\otimes \ (V^\star)^{\otimes n}: \quad\quad   & H^{VV^\star} & = \
H^V_L +
H_B+H^{V^\star}_R \\[2mm]
 V^{\otimes m} & \!\!\otimes\,  (V\otimes V^\star)^{\otimes L}
\otimes \ V^{\otimes n}: & H^{VV} & =  \ H^V_L + H_B+H^V_R \\[2mm]
 (V^\star)^{\otimes m} & \!\! \otimes \, (V\otimes
V^\star)^{\otimes L} \otimes \ (V^\star)^{\otimes n}:& H^{V^\star
V^\star} & = \ H^{V^\star}_L+ H_B +
H^{V^\star}_R\\[2mm]
 (V^\star)^{\otimes m} & \!\!\otimes \, (V\otimes V^\star)^{\otimes L}
\otimes \ V^{\otimes n}:& H^{V^\star V} & =  \ H^{V^\star}_L +
H_B+H^V_R,
\end{array}
\end{equation}
where the bulk Hamiltonian is the same as in
sec.~\ref{sec:untwisted} with $a=0$, i.e.\
\begin{equation}\label{eq:bulk_ham}
 H_B \ = \ - \sum_{i=m+1}^{2L+m-1}E_i - w
 \sum_{i=m+1}^{2L+m-2}P_{i,i+2}\ ,
\end{equation}
while the boundary Hamiltonians are as follows
\begin{align}\label{eq:b_hams_1}
 H^V_L \ =& -u\sum_{i=1}^m P_{i,i+1} \quad\quad\quad\quad\quad
 H^{V^\star}_R \ =\  -v \sum_{i=2L+m}^{2L+m+n-1} P_{i,i+1}\\ \label{eq:b_hams_2}
 H^{V^\star}_L \ =&   -u\sum_{i=1}^{m-1}P_{i,i+1} -  w' P_{m,m+2} -
 t'E_m  \\[2mm]
 H^V_R \ =  & -t''E_{2L+m} - w'' P_{2L+m-1,2L+m+1}-
 v \sum_{i=2L+m+1}^{2L+m+n-1}P_{i,i+1}\ .
 \end{align}
Taking into account that the monopole charges $M$ and $N$
describing the boundary conditions of the $\CPn$ sigma model can
be both positive and negative, the existence of four types of
chains~(\ref{eq:tw_chain_general}) labelled by two \emph{positive}
integers $m,n$ is quite suggestive of a possible identification.
On the other hand, the boundary conditions in the $\CPn$ sigma
model and the bundles associated to the corresponding branes do
not depend on the details of the connection, but only on their
curvature. The latter is essentially fixed by the monopole charge
$M$ or $N$. In view of the relation we are about to establish
between the spectrum of the $\CPn$ sigma model and that of the
chains~(\ref{eq:tw_chain_general}), the previous remarks raise the
question as to how much the spectrum of the
Hamiltonians~(\ref{eq:tw_chain_general}) depend on the precise
form of the boundary terms~(\ref{eq:b_hams_1}--\ref{eq:b_hams_2}).
We shall analyze this issue in the $\gl (1|1)$ subsector first.

\subsection{Numerics for the $\gl(1|1)$ subsector}\label{sec:gl11_subsector_tw}

To answer the question of universality and check the applicability
of formula~\eqref{eq:sm_tw_pred} to the
chains~(\ref{eq:tw_chain_general}), we first look at their
$\gl(1|1)$ subsectors. In this subsector, we can extend our
representation \eqref{eq:discrete_fermions} through discrete free
fermions to twisted open spin chain. With the boundary interaction
terms
\begin{align*}
P_{V\otimes~ V} &=\  -P_{V^\star\otimes~ V^\star} \  =\
[1 - (\bar{\varphi}_1-\bar{\varphi}_2)(\varphi_1-\varphi_2)]\\
P_{V\otimes~ V^\star\otimes~ V} &= \ -P_{V^\star\otimes V\otimes~
V^\star}\ = \
[1 -(\bar{\varphi}_1-\bar{\varphi}_3)(\varphi_1-\varphi_3)]\\
E_{V\otimes V^\star} &= \ - E_{V^\star \otimes V}\  =\
-(\bar{\varphi}_1-\bar{\varphi}_2)(\varphi_1+\varphi_2),
\end{align*}
we obtain a free system that can be studied numerically and with
great efficiency. Let us anticipate the following three basic
outcomes of the  numerical analysis.
\begin{enumerate}
\item
The $\gl(1|1)$ spin chains~(\ref{eq:tw_chain_general}) flow to the
free field theory of symplectic fermions with twisted boundary
conditions of the form \eqref{glue}.
\item
The twisting parameter $\lambda$ does not depend on the boundary
couplings $u,t',w',t''$, $w'',v$ as long as $t',t'',u$ and $v$ are
non-zero and the bulk length $L$ of the chain is sufficiently
large.
\item
In the continuum limit, the dependence of the twisting parameter $\lambda$ on $m,\, n$ and
$w$ for all four chains~\eqref{eq:tw_chain_general} is reproduced by eq.~\eqref{lform} for the $\CPn$ sigma model
with
\begin{equation}\label{eq:big_formula}
 \theta = \pi \ 
\end{equation}
provided the following identification between the monopole charges and the thickness of the boundaries of the chains is performed
\begin{align}\label{eq:mon_charges_bd_1}
 V^{\otimes m}\otimes (V\otimes V^\star)^{\otimes L}
\otimes (V^\star)^{\otimes n}:& &M\ =\ +m\quad N \ =\  +n \\[1mm]
\label{eq:mon_charges_bd_2}
 V^{\otimes m}\otimes (V\otimes V^\star)^{\otimes L}
\otimes V^{\otimes n}:& &M\ =\ +m\quad N\ =\ -n \\[1mm]
\label{eq:mon_charges_bd_3} (V^\star)^{\otimes m}\otimes
(V\otimes V^\star)^{\otimes L} \otimes (V^\star)^{\otimes
n}:& &M\ =\ -m\quad N \ =\  +n \\[1mm] \label{eq:mon_charges_bd_4}
 (V^\star)^{\otimes m}\otimes (V\otimes V^\star)^{\otimes L}
\otimes V^{\otimes n}:& &M\ = \ -m\quad N\ =\ -n\, .
\end{align}
\end{enumerate}
We now present the numerical evidence supporting these claims
one by one.

\begin{figure}[htbp]
\begin{center}
\includegraphics[scale=0.6]{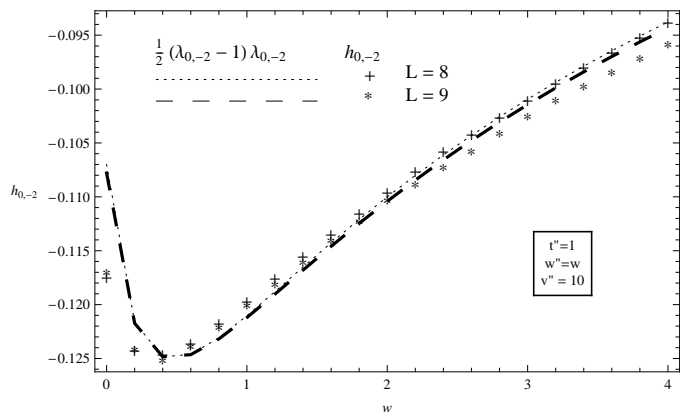}
\caption{Conformal dimension of the ground state of one of the
chains~(\ref{eq:tw_chain_general})
compared to the prediction provided by a twisted spectrum.}
\label{fig:proof_twisting}
\end{center}
\end{figure}

The numerical calculations supporting claim 1) are presented
in fig.~\ref{fig:proof_twisting}, where we compare the conformal
dimension $h$ for the ground state of our spin chain with the
expression
\begin{equation}
    h \ = \  \frac{\lambda(\lambda-1)}{2}\ \
\label{eq:tw_vac_formula}
\end{equation}
which determined the conformal dimension of twist fields in terms
of the twist parameter $\lambda$. For the lattice model, the twist
parameter is measured as the first excitation over the vacuum in
the $\gl(1|1)$ subsector.

\begin{figure}[htbp]
\begin{center}
\includegraphics[scale=0.6]{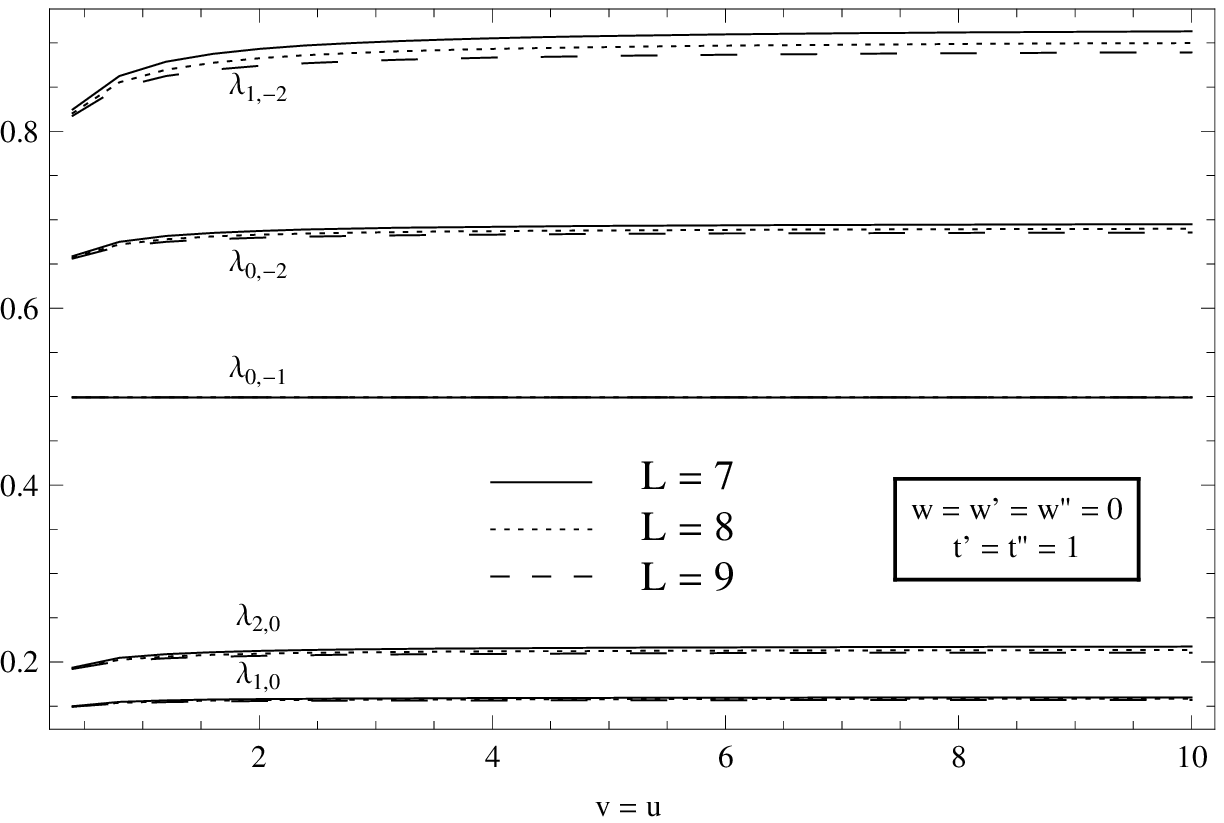}
\hfill
\includegraphics[scale=0.6]{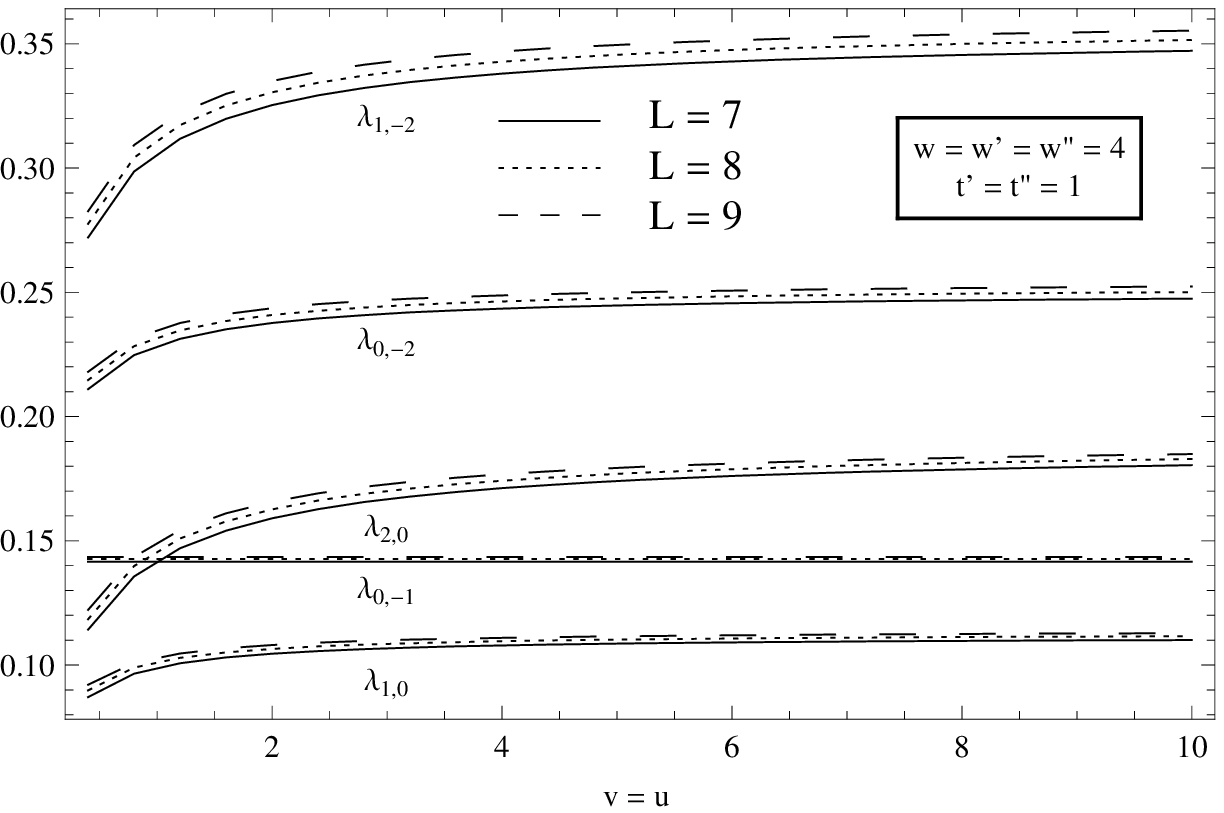}
\caption{Universality of $\lambda_{M,N}$ for several chains at
$w=0$ and $w=4$.} \label{fig:universality}
\end{center}
\end{figure}

Numerical evidence for the claim 2. on universality in the choice
of the boundary terms~(\ref{eq:b_hams_1}--\ref{eq:b_hams_2}) is
presented in fig.~\ref{fig:universality}.
Combining our claims 1. and 2. we see that for generic boundary
couplings $u,t',w',t'',w'',v$ the spectrum of the
Hamiltonian~(\ref{eq:tw_chain_general}), or at least of their
$\gl(1|1)$ subsectors, depend only on the thickness $m$ and $n$ of
the boundaries. In conclusion, the number of relevant parameters
in the four boundary terms~(\ref{eq:b_hams_1}--\ref{eq:b_hams_2})
exactly matches the number of parameters for the set of  boundary
conditions preserving the global symmetry of the $\CPn$ sigma
model.

\begin{figure}[htbp]
\begin{center}
\includegraphics[scale=0.6]{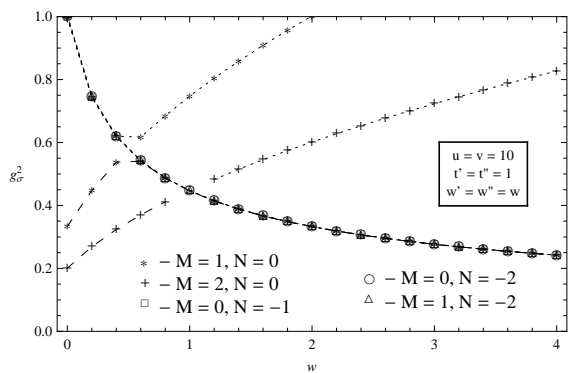}
\caption{Numerical evaluation of the one-to-one correspondence
between the $\CPn$ sigma model coupling constant $g_\sigma^2$ and
the bulk coupling constant $w$ of the spin
chains~(\ref{eq:tw_chain_general}). For the chains $N=0$ we
represent both branches for $g_\sigma^2$ expressed as a function
of $\tan \pi \lambda_{M,N}$. Calculations where made for $L=800$.}
\label{fig:gsigma2}
\end{center}
\end{figure}

Finally, we present in fig.~\ref{fig:gsigma2} compelling evidence
for the last claim 3. Using numerical data for $\lambda_{M,N}$ generated from
chains with different values of $M,\, N$ and $w$, we  plotted on
the same graph $g_\sigma^2$ expressed as a function of $\tan \pi
\lambda_{M,N}$ from eq.~\eqref{eq:sm_tw_pred} with $\theta = \pi$.
The appearance of a one to one correspondence between $w$ and
$g_\sigma^2$, which is independent of the chain we use, justifies
the applicability of eq.~\eqref{eq:sm_tw_pred} to the spin chains,
the correct value~\eqref{eq:big_formula} of the $\theta$-angle and the correct identification of the monopole
charge~(\ref{eq:mon_charges_bd_1}--\ref{eq:mon_charges_bd_4}).

This completes our analysis of the $\gl(1|1)$ subsector for the
chains~(\ref{eq:tw_chain_general}). So far, all our numerical
results were in perfect agreement with the continuum \CPzero\
sigma model. This supports our claim that the alternating
$\gl(N|N)$ spin chain provides a discretization for the $\CPn$
sigma model and it gives us sufficient confidence to address the
watermelon exponents for twisted spin chains with $S > 1$.
%

\subsection{Watermelon exponents for the twisted open chain}

Our aim now is to generalize the discussion of section
\ref{sec:untwisted} to the case of general monopole boundary
conditions. More precisely, we would like to determine the
conformal weight of tachyon vertex operators. For $l>0$, the latter are
associated with supersymmetric irreducible traceless tensors
$t(k+l,k)$ of contravariant rank $k+l$ and covariant rank $k$,
while for $l<0$ these are the supersymmetric irreducible traceless tensors
$t(k,k+|l|)$ of contravariant rank $k$ and covariant rank $k+|l|$.
In both case $k$ is a non-negative integer, which for
\CPone\ corresponds to the labels $\Lambda_{k,l}$ used before.

Let us restrict the algebraic
Hamiltonians~(\ref{eq:tw_chain_general}) to the representation of
the walled Brauer algebra provided by the space of embeddings of
the tensors $t(k+l,k)$ and $t(k,k+|l|)$ into the spin
chains~(\ref{eq:mon_charges_bd_1}--\ref{eq:mon_charges_bd_4}) with
monopole numbers $M$ and $N$. The lowest eigenvalue of the
Hamiltonian in each of these sectors will be called the
$(2k+|l|)$-legged watermelon exponent $h_{M,N}(k)$. As in the case
of the chain in sec.~\ref{sec:untwisted}, the watermelon exponents
all vanish in the limit $w\rightarrow 0$, i.e.\ in the region that
we associated with the large volume limit of the $\CPn$ sigma
model. The first two of these watermelon exponents are already
contained in the $\gl(1|1)$ subsector of the model, both in the
continuum and on the lattice. They are not degenerate. The
exponent $h_{M,N}(0)$ describes the twisted vacuum, while
$h_{M,N}(1)$ is associated with the first excitation. Their
non-zero difference is
\begin{equation}\label{eq:tw_wm_rel}
 \lambda_{M,N} \ = \ h_{M,N}(1)-h_{M,N}(0)\ .
\end{equation}

Another important observation coming from lattice calculations is
the Casimir evolution for the \emph{excitations} of the spin
chains~(\ref{eq:tw_chain_general}). Numerical calculations provide
compelling evidence that the following formula
\begin{equation}
    \delta h_{M,N}(k) \ = \ h_{M,N}(k) - h_{M,N}(0)  \ =\
    g_{M,N}\frac{k(k+|l|-1)}{2}\  \label{eq:cas_evolution_tw}
\end{equation}
holds for sufficiently large $w$ and with a universal function
$g_{M,N}$ that depends only on $M,N$ and $w$. In order to compare
with our continuum theory, we note that
\begin{equation}
\delta_l
C^{(2)}\left[\frac{l}{2}+k-1,0,\frac{l}{2}+2,\frac{l}{2}\right] -
\delta_l C^{(2)}\left[\frac{l}{2},0,
\frac{l}{2},\frac{l}{2}\right] \ = \ 2 k(k+l-1) \
\end{equation}
for $l = M-N > 0$. A similar result can be obtained when $l = M-N
< 0$. The expression $\delta_l C^{(2)}$ was defined in eq.\
\eqref{eq:deltaC}. The watermelon exponents $h_{M,N}(k)$ are
associated with the label $\Lambda_{k,l}$. The translation into
the label used in eq.\ \eqref{eq:deltaC} can be found at the end
of appendix B. In conclusion, we see that our lattice observation
\eqref{eq:cas_evolution_tw} for the watermelon exponents agrees
with their proposed continuum description in the \CPone\ model.

\begin{figure}%
\begin{center}
\includegraphics[scale=0.55]{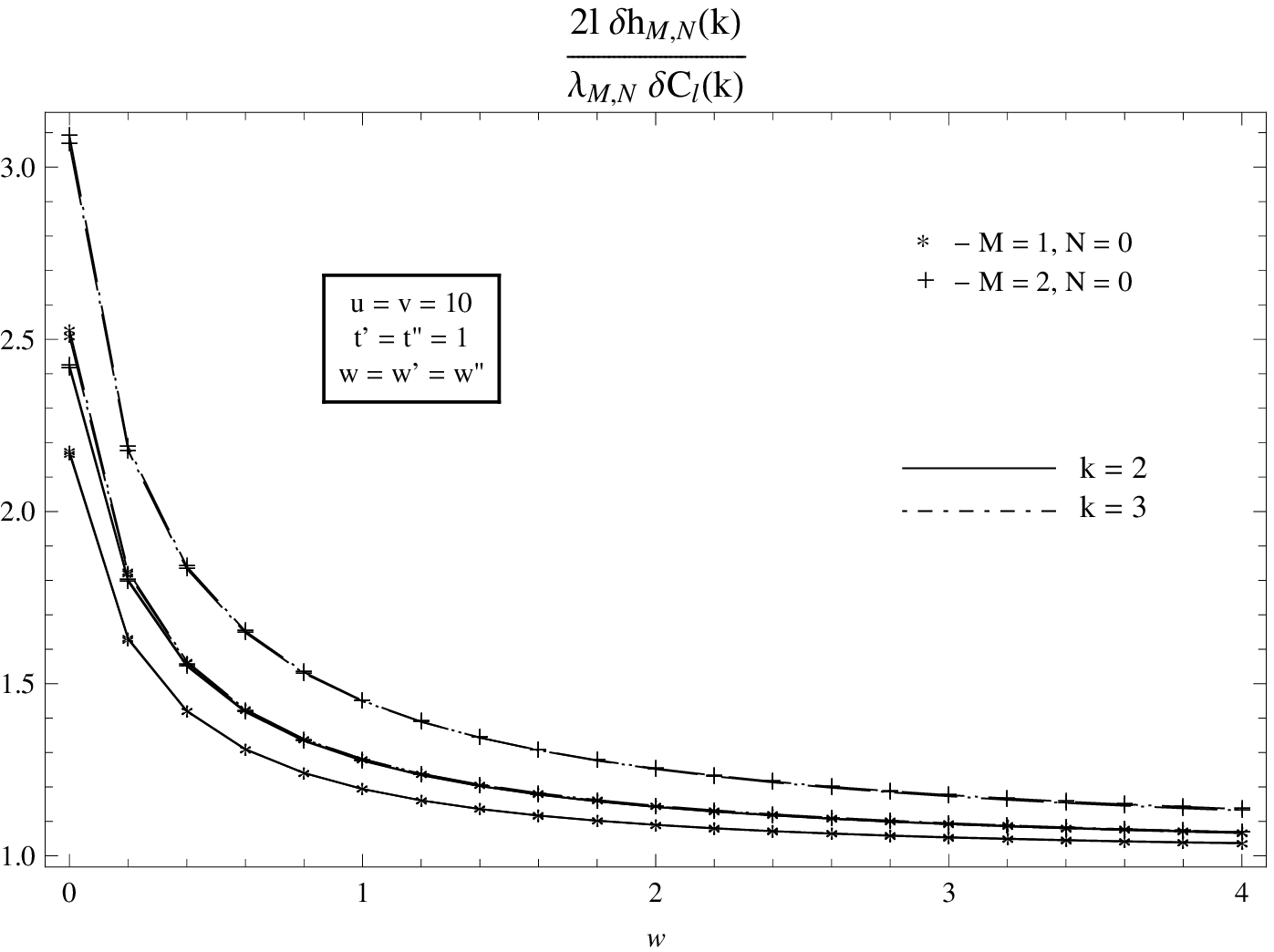}%
\hfill
\includegraphics[scale=0.55]{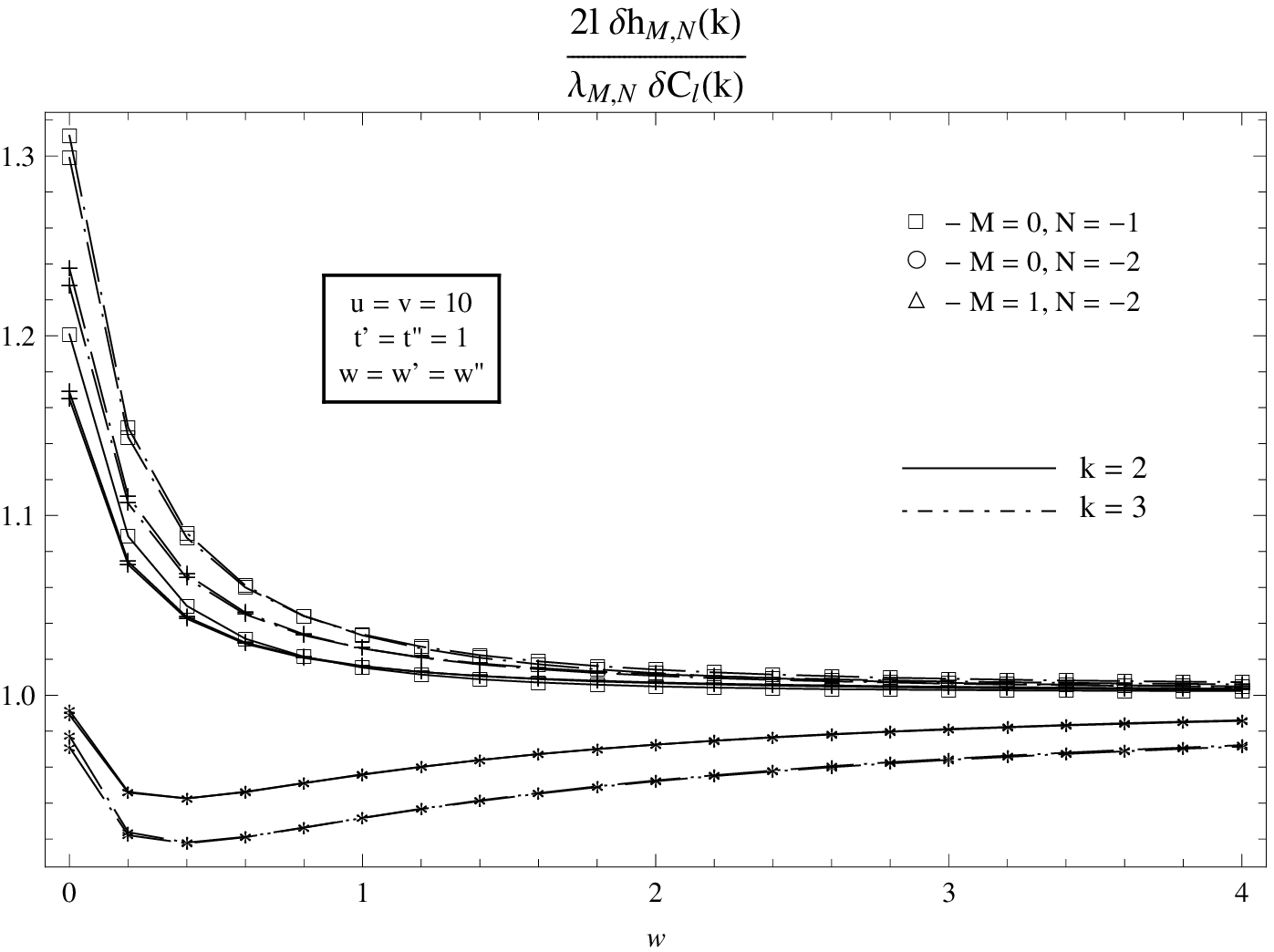}%
\end{center}
\caption{Test of eq.~\eqref{eq:cas_ev_constraint} following from
the assumption of Casimir evolution~\eqref{eq:cas_evolution_tw}.
Calculations where made for spin
chains~(\ref{eq:tw_chain_general})
of bulk length $L=7$ and $L=8$ and the corresponding curves
almost superpose.}%
\label{fig:ratio}%
\end{figure}

By analogy with sec.~\ref{sec:untwisted}, the function $g_{M,N}$
should be interpreted as the effective tension of the string
stretching between the bundle with monopole charge $M$ and the
bundle with monopole charge $N$. In the continuum theory, we
related the function $g_{M,N}$ to the twist parameter
$\lambda_{M,N}$ through the equation
\begin{equation}
    \lambda_{M,N} \ = \ \frac{|M-N|}{2} g_{M,N} \ .
\label{eq:cas_ev_constraint}
\end{equation}
It is interesting to test the validity of this relation
numerically. In fig.~\ref{fig:ratio}
we represent the ratio $|l| g_{M,N}/2\lambda_{M,N}$ as a function of
$w$. As before, we measure the function $g_{M,N}$ through the
equation \eqref{eq:cas_evolution_tw} for different excitations
$h_{M,N}(k)$. If the Casimir evolution~\eqref{eq:cas_evolution_tw}
holds true, then we should see a constant value of $|l| g_{M,N}/2\lambda_{M,N} =
1$ for the ratio, independently of the watermelon exponent that is
used to measure $g_{M,N}$. While things work out remarkably well
in the regime of large $w$, obvious discrepancies appear when $w$
is close to $w \sim  0$. The possible interpretation of these
differences are discussed in the next subsection.

\subsection{Comments on the region of small $w$}

There are actually several possibilities to interpret the failure
of  eq.~\eqref{eq:cas_ev_constraint} near $w=0$. We will discuss
two scenarios below. The ultimate test of the correct explanation
will be left for future work. In confronting our numerical results
with the proposed continuum description, we have tacitly assumed
that the spin chains~(\ref{eq:tw_chain_general}) at $w=0$ still
describe a point in the moduli space of the $\CPn$ sigma model.
This is a very strong assumption given that the symmetry of the
bulk Hamiltonian~\eqref{eq:bulk_ham} becomes much larger
\cite{SaleurRead} than $\gl(S|S)$ at $w=0$, essentially because the lines in the Brauer algebra representation are then prevented from crossing.

In assessing the meaning of the observed discrepancies, it is
useful to recall that a similar issue has also appeared for the
$\OSp(2S+2|2S)$ spin chain considered in~\cite{Candu:2008vw}. The
$\OSp$ spin chain was proposed as a discretization of the
$S^{2S+1|2S}$ supersphere sigma model. Generic features of the
lattice spectrum were found to be in excellent agreement with the
conjectured spectrum of the sigma model, as long as $w$ was large.
However, problems similar to the ones we described in the previous
subsection were encountered at the point $w=0$. Note that in the
supersphere case, the discrepancy was only visible when looking at
fields outside the $\OO(2)$ subsector of $\OSp(2S+2|2S)$ theory.
Again, a very similar observation was made for the $\gl (1|1)$
sector of the $\gl (S|S)$ spin chain. With all these similarities,
it seems likely that the discrepancies between lattice and
continuum analysis in the $\gl (S|S)$ and osp(2S+2$|$2S) model
should have the same explanation.

In the case of the supersphere sigma model, however, the
assumption of Casimir evolution for the whole spectrum stands on
rather firm grounds. To begin with, the perturbative expansion for
boundary conformal weights in the supersphere model may be summed
to all orders. Terms that could spoil the Casimir evolution were
shown to vanish. Moreover, world-sheet instanton corrections
cannot alter these findings, simply because they do not exist in
this case. Finally, a conjectured duality between the supersphere
sigma model and the osp(2S+2$|$2S) Gross-Neveu model was shown to
be perfectly consistent with the Casimir evolution of boundary
conformal weights \cite{Quella:2007sg, Mitev:2008yt}. All this 
makes it seem very likely that the conformal weights of the two 
investigated sigma models all evolve with the Casimir, as encoded 
in our formula
\eqref{ZRfinite}.

Having argued that the discrepancies between our lattice and
continuum results are unlikely to signal a breakdown of the
Casimir evolution in the sigma model, we want to entertain a
second logical possibility, namely that the continuum limit of the
spin chains~(\ref{eq:tw_chain_general}) is described by a $\CPn$
sigma model only for $w>0$, while at $w=0$ it is not. If this was
true then the discrepancies we observed in fig.~\ref{fig:ratio}
would simply result from interchanging the thermodynamic limit
$L\to\infty$ with the limit $w \to 0$. A similar non-commutativity
of limiting procedures can also be observed in the large volume
limit $w\to\infty$ where the symmetry of the Hamiltonian is once
more enhanced much beyond the generic $\gl (S|S)$ transformations.

Support for our second explanation of the discrepancies comes from
a closer inspection of the watermelon exponents. At $w=0$, the
lattice model is exactly solvable and we believe that the
differences between water-melon exponents are given by
\begin{equation}
\delta h_{M,N}(k)\ =\ \frac{k(k+2\lambda_{M,N}-1)}{2}\label{kac}
\end{equation}
where $\lambda_{M,N}$ is again measured as the difference
$\lambda_{M,N} = h_{M,N}(1) - h_{M,N}(0)$. The formula \eqref{kac}
can most certainly be derived analytically. But for now, we simply
justify it by observing that it fits the general pattern of
boundary exponents in (non intersecting) loop models discussed in
\cite{JS}. Indeed, it can be rewritten as
\begin{equation*}
h_{M,N}(k)\ =\ h_{2\lambda_{M,N}-1,2\lambda_{M,N}-1+2k}
\end{equation*}
where on the right hand side we use the Kac formula at $c=-2$:
\begin{equation*}
h_{r,s}=\frac{(2r-s)^2-1}{8}
\end{equation*}
A verification of this formula is presented in tab.~\ref{guess}.
The numbers in the grid should all go to unity in the scaling
limit. We see that the agreement with eq.~\eqref{kac} is quite
impressive. The behavior of watermelon exponents in the chain with
$w\neq 0$ is significantly different. This supports our claim that
the continuum theory of the $w=0$ lattice model does not belong to
the continuous family of conformal field theories that is
parametrized by $w >0$.

\begin{table}
\begin{center}
  \begin{tabular}{|c|c|c|c|c|}
     \hline
     \multirow{2}{*}{$M$} & \multirow{2}{*}{$N$} &
        \multicolumn{3}{|c|}{$\tfrac{2\delta
h_{M,N}(k)}{k(k+2\lambda_{M,N}-1)}$}\\ \cline{3-5}
        & & $k=2$ & $k=3$ & $k=4$ \\ \hline
    \input{table.dat} \\
    \hline
  \end{tabular}
\end{center}
  \caption{Numerical check of the proposed formula~\eqref{kac} for
  the watermelon exponents of the spin chains~(\ref{eq:tw_chain_general})
  at $w=0$. Calculations where made
  for bulk length $L=7$.}
  \label{guess}
\end{table}
%

\section{Conclusions and Open Problems}

In this work we have analyzed the boundary partition functions for
all \utwo\ invariant boundary conditions of the sigma model in the
projective superspace \CPone. The dependence of this partition
function on the bulk couplings $g_\sigma$ and $\theta$ and on the
boundary monopole charges $M,N$ was displayed in eq.\
\eqref{ZRfinite}. It contains the branching functions
\eqref{branchingKac1} of the model at $R=\infty$ along with two 
universal functions $\lambda_{M,N}$ and $g_{M,N}$ which are
defined through eqs.\ \eqref{lform} and \eqref{flrel},
respectively. The partition function encodes the dependence of
boundary conformal weights on the various couplings
and justifies and generalizes the results in \cite{Candu}.
In the second
part, we introduced a lattice model with Hamiltonian
\eqref{eq:general_ham} on an alternating spin chain. Numerical
studies of the latter revealed an excellent agreement with the
predictions from the continuum theory, at least for sufficiently
large values of the lattice coupling $w$. In particular, we were
able to model all the boundary conditions of the continuum theory
by adding boundary layers of finite width to the open spin chain,
see eq.\ \eqref{eq:tw_chain_general}.

One of the most interesting applications of our results would be
to search for values of the parameters $g_\sigma$ and $\theta$ at
which the world-sheet symmetry gets enhanced, e.g.\ to some affine
Lie algebra symmetry. A similar Wess-Zumino point exists for sigma
models on superspheres $S^{2S+1|2S}$ and it gives rise to an
interesting dual description of the theory through a non-geometric
Gross-Neveu model. It is very likely that similar points exist for
sigma model on complex projective superspaces as well. Even though
we have not yet been able to identify a point with affine
psl(2$|$2) symmetry, we hope to return to this issue soon.

Another possible further direction concerns the closely related
non-compact sigma model on the coset space
$\ensuremath{\text{U}(1,1|2)}/\ensuremath{\text{U}(1|1)}\times
\ensuremath{\text{U}(1|1)}$ that was considered in
\cite{Zirnbauer} because of its possible relevance for the theory
of quantum Hall plateau transitions. The spin chain discussed in
\cite{Zirnbauer} involves infinite dimensional representations and
a pure Heisenberg interaction.\footnote{
This chain was proposed earlier in unpublished work by N. Read.
}
It would be interesting, among
other things, to study the role of next to nearest neighbor
interactions in that case, and to analyze whether they allow fine
tuning of the running coupling constant as in our model. It could
also be of interest to interpret our bundle boundary conditions in
terms of edge states in the Hall effect \cite{Xiong}.

A striking conclusion of our study is that, like in the
supersphere case, the chain with the simplest interaction (no loop
crossing in the Brauer formulation, or $w=0$) seems to be in a
different universality class from the generic $w\neq 0$ case.
Non-commutativity of the limits $L\to\infty$ and $w\to 0$ means
more precisely that the perturbation induced by turning  $w\neq 0$
on the lattice is \emph{relevant} at $w=0$. The conformal field
theory at that point admits a very large symmetry, but  has not
yet been fully explored. For the whole picture to be consistent,
the \emph{bulk} spectrum should contain  an invariant, marginally
relevant operator, which should moreover be absent in the  minimal
$\U{1}{1}$ or $\OO(2)$  subsector. The existence of this operator
remains to be established.

Let us point out that there are some other predictions of the Casimir evolution that could be checked in the large volume regime.
Note that the Casimir evolution for the weights of tachyonic vertex operators is supported by both perturbative and non-perturbative numerical calculations only in the theory with equal boundary monopole charges $M=N$.
While the conjectured exact form~\eqref{last_eq_label} of watermelon exponents in the theory with arbitrary boundary monopole charges $M,N$ passed several analytical and numerical tests,
it could not be backed up by perturbative calculations beyond
the leading order because we did not succeed to generalize the background field expansion to twisted boundary conditions.
Nonetheless, we suspect that such a generalization exists and the
watermelon exponents will most likely be computed again
in terms of eigenvalues of some Laplacian on the bundle with monopole charge $l=M-N$.
The point is that for $l\neq 0$ this Laplacian is not unique, 
as can be seen from the existence of a 1-parameter family of $\gl(S|S)$ Casimirs $\Cas_\alpha$, see app.~\ref{sec:Laplacian}.
However, if we choose
\begin{equation*}
  \alpha=1-\frac{g_{M,N}(g_\sigma,\theta)}{2}
\end{equation*}
then the conjectured form~\eqref{last_eq_label}
for the watermelon exponents coincides exactly with  a Casimir evolution type formula 
\begin{equation*}
 h^{g_\sigma,\theta}_{M,N}(k) = \frac{g_{M,N}(g_\sigma,\theta)}{4}\Cas_{\alpha}(\Lambda_{k,l}),
\end{equation*}
which is most natural in the context of the background field method.
On the other hand, these conjectured watermelon
exponents possess the following simple expansion in the coupling
$g_\sigma$,
\begin{equation*}
  h_{M,N}(k) \ = \ \frac{g^2_\sigma}{\pi}\Cas_{\alpha=1}(\Lambda_{k,l})+
  \frac{2g_\sigma^4}{\pi^2}
l^2+O(g_\sigma^6)\ .
\end{equation*}
Here, $\Cas_{\alpha=1}(\Lambda_{k,l})$ are the eigenvalues of the
Bochner-Laplacian of the complex line bundles over $\CPn$ and, as we said, the first term can be reproduced by the semi-classical approximation.
In the case $l \neq 0$ the first correction to the
semi-classical result comes at order $g_\sigma^4$.
This is an accessible non-trivial check to be performed once the perturbation theory for twisted boundary conditions is ironed out.

Moving away from $\theta=\pi$ in the sigma model corresponds to
staggering the couplings of the spin chain. In the case $w=0$, it
is well known that staggering in fact does not affect the spectrum
at all. For $w\neq 0$, we expect in general that staggering will
renormalize the coupling constant to which the lattice model flows
(so the $g_\sigma^2(w)$ dependence will be now a dependence on $w$
and the staggering parameter), on top of affecting the value of
$\theta$ in the formulas. Our continuum theory makes rather
non-trivial predictions about this functional dependence that seem
well worth further investigation.
\medskip

{\bf Acknowledgments:}
We especially thank N. Read for an earlier collaboration on the subject, and for many illuminating comments and discussions.
We thank Nathan Berkovits, Thomas
Creutzig, Manfred Herbst, Marcos Marino, Tristan McLoughlin,
Nikita Nekrasov, Nick Read, Soo-Jong Rey, Peter R\o nne, Sakura
Sch\"afer-Nameki and Edward Witten for discussions and comments. We are
gratefull to the Galileo Galilei Institute for its hospitality during the
beginning of this work. 
This research was supported in part by the National Science
Foundation under Grant No. PHY05-51164. The research of TQ is funded by
a Marie Curie Intra-European Fellowship, contract number MEIF-CT-2007-041765. HS was supported by the ANR and the Network INSTANS.

\begin{appendix}

\section{The quadratic Casimir elements}

\label{sec:most_gen_2_cas}

For a simple contragredient Lie superalgebra $\mathfrak{g}$ the
invariant, supersymmetric, consistent, non-degenerate and bilinear
form $\beta:\mathfrak{g}\times \mathfrak{g}\rightarrow \mathbb{C}$
exists and is defined uniquely up to a proportionality constant.
Every such invariant form $\beta$ defines a quadratic central
element in the universal enveloping Lie superalgebra in the
standard way. To be more precise, let $T_a$ be a basis of
$\mathfrak{g}$ and let $T^a$ be the dual basis with respect to
$\beta$, that is
\begin{equation}
    \beta(T^a,T_b)\  = \ \delta^a_b.
\label{eq:dual_basis}
\end{equation}
Then the quadratic Casimir associated to the invariant form
$\beta$ is defined as
\begin{equation}
    \Cas \ = \ \sum_a T_a T^a \ .
\label{eq:def_cas}
\end{equation}
It is not hard to verify that $\Cas$ is indeed central. The Lie
superalgebra $\u$ we are dealing with in this work, however, is
not simple. After a normalization has been fixed, it possesses a
one parameter family of invariant, supersymmetric, consistent,
non-degenerate and bilinear forms. Let $V\simeq V_{\bar{0}}\oplus
V_{\bar{1}}$ denote the graded fundamental module of $\u$ with
even dimension $\dim V_{\bar{0}}=S$ and odd dimension $\dim
V_{\bar{1}} = S$ and $R_V : \u\rightarrow \gl(V)$ be the
corresponding representation. Then the one parameter space of
invariant forms of $\u$ is constructed by using the invariant
supertrace
\begin{equation}
    \beta(X,Y) \ =\  \str R_V (X Y)  +  \alpha \str R_V(X) \str R_V(Y)\ .
\label{eq:two_par_fam_in_forms}
\end{equation}
Let now $E_{i\phantom{j}}^{\phantom{i}j}$ be the standard basis of
$\gl(V)$, that is the $2S\times 2S$ matrices with an entry 1 in
the $i$-th row and $j$-th column and 0 entries everywhere else.
According to the def.~\eqref{eq:dual_basis}, the basis dual to
$E_{i\phantom{j}}^{\phantom{i}j}$ with respect to the
form~\eqref{eq:two_par_fam_in_forms} is given by
\begin{equation*}
    \Big(E_{i\phantom{j}}^{\phantom{i}j}\Big)^* \ = \
    (-1)^{|j|}E_{j\phantom{i}}^{\phantom{j}i} - \alpha \delta^i_j E\ ,
\end{equation*}
where we have denoted by $E$ the identity matrix. The quadratic
Casimir of a reductive Lie superalgebra is constructed in the same
way as in eq.\ \eqref{eq:def_cas}. When the invariant forms are
not unique, the same is true for the Casimir element. In
particular, the quadratic Casimir element of $\u$ that is
associated to the form~\eqref{eq:two_par_fam_in_forms} becomes
\begin{equation}
    \Cas_\alpha \ = \ E_{i\phantom{j}}^{\phantom{i}j}
    E_{j\phantom{i}}^{\phantom{j}i}(-1)^{|j|} - \alpha E^2\ .
\label{eq:explicit_glnn_cas}
\end{equation}
The eigenvalues of $\Cas_\alpha$ in an irreducible representation
with highest weight $\Lambda$ can be evaluated by computing scalar
products in the weight space $\mathfrak{h}^*$ in exactly the same
way as for simple Lie superalgebras.  Let us see how this works.
Choose the diagonal generators $D_1 =
E_{1\phantom{1}}^{\phantom{1}1},\dots,D_{2S}=E_{2S\phantom{2S}}^{\phantom{2S}2S}
$
as a basis of the Cartan subalgebra $\mathfrak{h}$ of $\u$ and
denote by $\epsilon_1,\dots,\epsilon_S,\delta_1, \dots, \delta_S$,
respectively, the dual basis in $\mathfrak{h}$. The restriction of
$\beta$ to $\mathfrak{h}$ defines a natural isomorphism
$\varphi:\mathfrak{h}\rightarrow \mathfrak{h}^*$ by
\begin{equation}
    \varphi(H')(H'') \ = \ \beta(H',H'')
\label{eq:nat_isom}
\end{equation}
and endows $\mathfrak{h}^*$ with a scalar product
\begin{equation}
    (\lambda,\mu)_\alpha \ = \ \beta \big(\varphi^{-1}
    (\lambda),\varphi^{-1}(\mu)\big)\ .
\label{eq:scalar_prod_weight_space}
\end{equation}
In the basis $\delta_i,\epsilon_j$ of $\mathfrak{h}^*$, the
natural isomorphism~\eqref{eq:nat_isom} reduces to
\begin{equation*}
    \varphi(D_1) \ = \ \epsilon_1,\dots,
    \varphi(D_{2S}) \ = \ \delta_{S}.
\end{equation*}
The matrix elements of the scalar
product~\eqref{eq:scalar_prod_weight_space} in the weight space
$\mathfrak{h}^*$ of $\u$ with respect to the basis
$\epsilon_i,\delta_j$ can easily be computed
\begin{equation}
    (\epsilon_i,\epsilon_j)_\alpha \ = \ \delta_{ij} -
    \alpha,\quad (\delta_i, \delta_j)_\alpha
    = -\delta_{ij} - \alpha,\quad
    (\epsilon_i,\delta_j)_\alpha = -\alpha\ .
\label{eq:mat_el_scalar_prod}
\end{equation}
One natural way to parametrize the highest weight vectors
$\Lambda$ for irreducible representations of $\u$ is by specifying
the coordinates of $\Lambda$ with respect to the basis
$\epsilon_i,\delta_j$. Thus, if
\begin{equation}
    \Lambda \ = \ \sum_{i=1}^S \rho_i \delta_i + \sigma_i \epsilon_i
\label{eq:coordinates_yt_hw}
\end{equation}
is the highest weight of a highest weight representation, then
\begin{equation}
    \sigma_i \ = \ \Lambda(D_i), \quad \rho_i \ = \
    \Lambda(D_{S+i}),\qquad i = 1,\dots,S\ .
\label{eq:some_trivialities}
\end{equation}
The eigenvalues of the Casimir element do not depend on the
conventions for positiveness in the weight space. To compute them,
we shall use a non-standard, but convenient absolute ordering
\begin{equation}
    \epsilon_1 \ > \ \dots \ >\  \epsilon_S\  > \ \delta_1\  >\
     \dots\  > \ \delta_S
\label{eq:convenient_choice_abs_ord}
\end{equation}
which fixes the positive roots to
\begin{equation*}
    \epsilon_i - \epsilon_j, \quad \delta_k - \delta_l,\quad
    \epsilon_i - \delta_k\ ,
\end{equation*}
where $i<j$ and $k<l$. Now if $v_\Lambda$ is the highest weight
vector of some highest weight representation, then the eigenvalue
of the Casimir on that representation can be easily computed
\begin{align*}
    \Cas_\alpha v_\Lambda &=\  \sum_{i=1}^{2S} (-1)^{|i|}D_i^2v_\Lambda
    - \alpha E^2 v_\Lambda +
    \sum_{j=2}^{2S}\sum_{i=1}^{j-1}[E_{i\phantom{j}}^{\phantom{i}j},
    E_{j\phantom{i}}^{\phantom{j}i}](-1)^{|j|}
    v_\Lambda \\[2mm]
    &=
    \left(\sum_{i=1}^{2S} (-1)^{|i|}\Lambda(D_i)^2 - \alpha \Lambda(E)^2 +
    \sum_{j=2}^{2S}\sum_{i=1}^{j-1}[(-1)^{|j|}\Lambda(D_i) -
    (-1)^{|i|}\Lambda(D_j)]\right) v_\Lambda\ .
\end{align*}
Using the eqs.\ (\ref{eq:mat_el_scalar_prod},
\ref{eq:coordinates_yt_hw} and \ref{eq:some_trivialities}) one can
derive the desired form for the eigenvalues $\Cas_\alpha(\Lambda)$
of the Casimir~\eqref{eq:explicit_glnn_cas} in a highest weight
representation with highest weight $\Lambda$, namely
\begin{equation}
    \Cas_\alpha(\Lambda) \ = \ (\Lambda,\Lambda + 2\rho)_\alpha,
\label{eq:formula_for_cas_egs}
\end{equation}
where $\rho$ is the Weyl vector
\begin{equation*}
    2\rho \ = \ \sum_{1\leq i<j\leq S} (\epsilon_i -\epsilon_j +
    \delta_i - \delta_j) -
    \sum_{i,j=1}^S(\epsilon_i - \delta_j)
\end{equation*}
with respect to the chosen absolute
ordering~\eqref{eq:convenient_choice_abs_ord}. Keeping in mind
that the Weyl vector is the half sum of all positive roots minus
the half sum of all negative roots, the formula
eq.~\eqref{eq:formula_for_cas_egs} for the eigenvalues of the
Casimir can be rendered independent of the definition of
positiveness in the weight space.

In the paper we use another notation for the weights of \utwo,
which stems from a different choice \eqref{Cartanutwo} of basis
for the Cartan algebra. With respect to this basis, a  highest
weight $\Lambda=[j_1,j_2,a,b]$ has the following components
\begin{equation}
\Lambda(J_x) \ = \ j_1\, , \quad \Lambda(J_y)\  = \ j_2,\quad
\Lambda(J_z) \ = \ a\, ,\quad \Lambda(J_u) \ = \ b\ .
\label{eq:dynkin_labels}
\end{equation}
The dictionary between the labels $\rho_i,\sigma_j$ of
eq.~\eqref{eq:coordinates_yt_hw} and the labels $j_1,j_2,a,b$ is
easy to establish
\begin{equation}
    \sigma_1 -\sigma_2 \ = \ 2j_1\, ,\quad  \rho_1 - \rho_2 \ = \ 2j_2\, ,
    \quad \sigma_1+\sigma_2-\rho_1-\rho_2 \ = \ 2a\, ,\quad
    \sigma_1+\sigma_2 + \rho_1+\rho_2\  = \ 2b\ .
\label{eq:dict}
\end{equation}
Moreover, from eq.~\eqref{eq:formula_for_cas_egs} we obtain our
formula \eqref{eq:Casimiralpha} for the value of the Casimir
elements in the representations $[j_1,j_2,a,b]$ of \utwo.

\section{Laplacian on complex line bundles over \CPn}

\label{sec:Laplacian}

Let $g_{pq}$ be the matrix elements of the metric $g$ on \CPn\ in
some set of local real coordinates $\varphi^p$, $g^{pq}$ be the
matrix inverse to $g_{pq}$, $\nabla$ be the Levi-Civita connection
with respect to the metric $g$ and $A = A_p(\varphi) d \varphi^p$
be the one-form monopole defining a complex line bundle over \CPn.
Then the Bochner-Laplacian on the complex line bundle over \CPn\
is defined by the following second order, $\u$-invariant
differential operator
\begin{equation*}
    \Delta = g^{pq} (\nabla_p + A_q)(\nabla_p + A_q).
\end{equation*}

Existence theorems~\cite{Konstant} ensure that a non-trivial
complex line bundle exists and is unique if and only if the
curvature  $\Omega = dA$ of the connection $A$ satisfies the
following integrality condition
\begin{equation*}
    \int_{\bCPone} \frac{\Omega}{2\pi i} \in \mathbb{Z}.
\end{equation*}

Let $w^i$ be a set of local holomorphic coordinates on \CPn. Then
the standard metric on  \CPn\ is given by the Fubini-Study metric
\begin{equation*}
    g_{i\bar{\jmath}} = \frac{\delta_{ij}}{1+w^\dagger\cdot w}-
    \frac{(-1)^{|j|}w^{\bar{\imath}}w^j}{(1+w^\dagger \cdot w)^2},
\end{equation*}
where the sign conventions for the scalar product in the
supereuclidean space $\mathbb{C}^{S-1|S}$ are $w^\dagger\cdot
w=\delta_{ij}w^{\bar{\jmath}} w^i $. The metric form is
\begin{equation*}
    ds^2 = g_{pq}d\varphi^p d\varphi^q = 2
g_{i\bar{\jmath}}dw^{\bar{\jmath}}dw^i
\end{equation*}
and all the geodesics are closed and of fixed length $\sqrt{2}\pi$. The
K\"{a}hler form
\begin{equation*}
    \omega = -i g_{i\bar{\jmath}} dw^{\bar{\jmath}}\wedge dw^i
\end{equation*}
can be normalized to yield a generator for the second integral
cohomology group. Indeed, from
\begin{equation*}
    \int_{\bCPone} \omega = 2\pi,
\end{equation*}
the existence condition for the complex line bundle reduces to
\begin{equation*}
    \Omega = -i l \omega,
\end{equation*}
where $l\in\mathbb{Z}$ is called the monopole charge.

By standard methods in the theory of complex line bundles,
see~\cite{kuwabara}, one can prove that the space of sections of
the bundle with monopole charge $l$ is isomorphic to the space of
equivariant functions on \CPn, that is the space of functions
$f(w,\bar{w})$ with the property
\begin{equation*}
    f(e^{i\alpha} w, e^{-i\alpha} \bar{w}) = e^{i\alpha l} f(w,\bar{w}),
\end{equation*}
where $\alpha$ is real. This functional space
can be constructed as a square integrable  span on the monomials
$Z^{i_1}\dots Z^{i_{k+l}}\bar{Z}^{j_1} \dots\bar{Z}^{j_k}$, where
$Z^i$ are the components of some vector belonging to the
$\u$-fundamental representation $\square$ satisfying $Z^\dagger \cdot Z
= 1$ and $k,l$ are integers such that $k\geq 0,\, k+l\geq 0$.

The harmonic decomposition of the space of equivariant functions
with monopole charge $l\neq 0$ is a multiplicity free direct sum of
$\u$ supersymmetric traceless irreducible tensors $t(k+l,k)$ of
contravariant rank $k+l\geq 0$ and covariant rank $k\geq0$.
The highest weights of these tensors can be easily computed in the
$\delta_i,\epsilon_j$ basis of sec.~\ref{sec:most_gen_2_cas}. If
one chooses the absolute
ordering~\eqref{eq:convenient_choice_abs_ord} in the weight space
of $\u$ then the highest weight of the fundamental representation
becomes $\epsilon_1$, while of that of the dual representation
$-\delta_S$. The weight of a supersymmetric tensor power of a
vector follows immediately from the definition of the tensor
action of the superalgebra. Thus, the highest weights of the
supersymmetric irreducible traceless tensors $t(k+l,k),\, l>0$ are
\begin{equation*}
    \Lambda_{k,l} = \begin{cases}
    (k+l)\epsilon_1  -   \delta_{S-k+1} - \dots - \delta_S, & k\leq S\\
    (k+l)\epsilon_1 -(k-S)\epsilon_S -\delta_1 - \cdots - \delta_S, & k> S
    \end{cases},
\end{equation*}
while those of the tensors $t(k'+l,k') = t(k,k+|l|),\, l<0$ are
\begin{equation*}
    \Lambda_{k,l} = \begin{cases}
    k\epsilon_1 - \delta_{S-k-|l|+1}-\dots -\delta_S, & k+|l| \leq S \\
    k\epsilon_1 - (k+|l|-S)\epsilon_1 - \delta_1-\dots-\delta_S, & k+|l|> S
    \end{cases},
\end{equation*}
where in both cases $k\geq 0$.

With this explicit construction of the complex line bundles at
hand one can compute the spectrum of the Bochner-Laplacian, see
~\cite{kuwabara}. The net result for the eigenvalues
$\lambda_l(k)$ of $\Delta$ is
\begin{equation}
    \lambda_{l}(k) = -2\left(k + \frac{|l|}{2}\right)\left(k + \frac{|l|}{2} - 1
\right) + \frac{l^2}{2},
\label{eq:}
\end{equation}
where $k\geq 0$.
Comparing this spectrum to the eigenvalues of the
Casimir~(\ref{eq:explicit_glnn_cas}, \ref{eq:formula_for_cas_egs})
\begin{equation}
    \Cas_\alpha(\Lambda_{k,l}) = 2k^2 + (2k+|l|)(|l|-1) - \alpha l^2,
\label{eq:cas_eg_standard}
\end{equation}
we see that
\begin{equation*}
    \Delta = -\Cas_{\alpha=1}.
\end{equation*}

In the end let us list the labels~\eqref{eq:dynkin_labels}  of the
highest weights $\Lambda_{k,l}$ of supersymmetric traceless
irreducible \utwo-tensors $t(k+l,k)$ and $t(k,k+|l|)$. Using the
dictionary~\eqref{eq:dict} we get for $l\geq 0$
\begin{equation*}
    \Lambda_{0,l} = \left [ \frac{l}{2}, 0, \frac{l}{2}, \frac{l}{2}
\right],\quad
    \Lambda_{1,l} = \left [ \frac{l+1}{2}, \frac{1}{2}, \frac{l}{2}+1,
\frac{l}{2} \right],\quad
    \Lambda_{k,l} = \left [ \frac{l}{2}+k-1, 0, \frac{l}{2}+2, \frac{l}{2}
\right],
\end{equation*}
for $k=2,3,\dots$. When $l<0$ we have
\begin{equation*}
    \Lambda_{0,-1} = \left[ 0,\frac{1}{2}, \frac{1}{2},-\frac{1}{2}
\right],\quad
    \Lambda_{k,l} = \left[  -\frac{l}{2}+k-1, 0, \frac{l}{2}+2, \frac{l}{2}
\right],\quad k+|l|\geq 2.
\end{equation*}

\section{Atypical branching functions}

\label{sec:atypicalBF}

In this appendix we collect explicit formulas for the branching
functions of atypical \utwo\ representations in terms of those for
Kac-modules. Let latter were displayed in the main text. As in the
rest of the paper, finite dimensional representations of \utwo\
are labelled by four parameters $j_1,j_2 \in \mathbb{N}/2$ and
$a,b \in \mathbb{R}$. There are five different kinds of
atypicality conditions on these labels. For each of these we shall
then list the atypical branching functions. All of them can be
derived using the character formulas in \cite{ZhangGould}.
\begin{itemize}
\item{$b=j_1-j_2=0$}
\beqa
\psi_{[0,0,a,0]}&=&\psi_{[0,0,a,0]}^K
+\psi_{[0,0,a+4,0]}^K+\psi_{\left[\frac{1}{2},\frac{1}{2},a+1,0\right]}^K+\psi_{
\left[ \frac { 1 } { 2
},\frac{1}{2},a+3,0\right] } ^K\nonumber\\[2mm]
\psi_{\left[\frac{1}{2},\frac{1}{2},a,0\right]}&=&\psi_{\left[\frac{1}{2},\frac{
1 } { 2 } , a , 0 \right] } ^K
+\psi_{\left[\frac{1}{2},\frac{1}{2},a+2,0\right]}^K+\psi_{[0,0,a+1,0]}^K+\psi_{
[ 1 , 1 , a+1 , 0 ]
} ^K\\[2mm]
\psi_{[j,j,a,0]}&=&\psi_{[j,j,a,0]}^K
+\psi_{[j,j,a+2,0]}^K+\psi_{\left[j-\frac{1}{2},j-\frac{1}{2},a+1,0\right]}
^K+\psi_ { \left[
j+\frac{1}{2}, j+\frac{1}{2},a+1,0\right]} ^K\text{ for } j\geq 1
\nonumber\eeqa
\item{$b=j_1-j_2\neq0$}
\beqa
\psi_{\left[\frac{1}{2},0,a,\frac{1}{2}\right]}&=&\psi^K_{\left[\frac{1}{2},0,a,
\frac{1}{2}\right ] }
+\psi^K_{\left[0,\frac{1}{2},a+3,\frac{1}{2}\right]}\nonumber\\[2mm]
\psi_{\left[0,\frac{1}{2},a,-\frac{1}{2}\right]}&=&\psi^K_{\left[0,\frac{1}{2},0
, a ,
-\frac{1}{2}\right ] }
+\psi^K_{\left[\frac{1}{2},0,a+3,-\frac{1}{2}\right]}\nonumber\\[2mm]
\psi_{\left[j_1,0,a,j_1\right]}&=&\psi^K_{\left[j_1,0,a,j_1\right]}
+\psi^K_{\left[j_1-1,0,a+2,j_1\right]}\text{ for } j_1\geq 1
\\[2mm]
\psi_{\left[0,j_2,a,-j_2\right]}&=&\psi^K_{\left[0,j_2,a,-j_2\right]}
+\psi^K_{\left[0,j_2-1,a+2,-j_2\right]}\text{ for } j_2\geq
1\nonumber\\[2mm]
\psi_{\left[j_1,j_2,a,j_1-j_2\right]}&=&\psi^K_{\left[j_1,j_2,a,j_1-j_2\right]}
+\psi^K_{\left[j_1-\frac{1}{2},j_2-\frac{1}{2},a+1,j_1-j_2\right]}\text{ for }
j_1\text{ and } j_2\geq 0\nonumber
\eeqa
\item{$b=-j_1+j_2\neq0$}
\beqa
\psi_{\left[j_1,j_2,a,-j_1+j_2\right]}&=&\psi^K_{\left[j_1,j_2,a,
-j_1+j_2\right ] }
+\psi^K_{\left[j_1+\frac{1}{2},j_2+\frac{1}{2},a+1,-j_1+j_2\right]}
\eeqa
\item{$b=j_1+j_2+1$}
\beqa
\psi_{\left[0,j_2,a,j_2+1\right]}&=&\psi^K_{\left[0,j_2,a,
j_2+1\right ] }
+\psi^K_{\left[0,j_2+1,a+2,j_2+1\right]}\\[2mm]
\psi_{\left[j_1,j_2,a,j_1+j_2+1\right]}&=&\psi^K_{\left[j_1,j_2,a,
j_1+j_2+1\right ] }
+\psi^K_{\left[j_1-\frac{1}{2},j_2+\frac{1}{2},a+1,j_1+j_2+1\right]}
\text{ for } j_1\geq \frac{1}{2}\nonumber
\eeqa
\item{$b=-j_1-j_2-1$}
\beqa
\psi_{\left[j_1,0,a,-j_1-1\right]}&=&\psi^K_{\left[j_1,0,a,
-j_1-1\right ] }
+\psi^K_{\left[j_1-1,0,a+2,-j_1-1\right]}\\[2mm]
\psi_{\left[j_1,j_2,a,-j_1-j_2-1\right]}&=&\psi^K_{\left[j_1,j_2,a,
-j_1-j_2-1\right ] }
+\psi^K_{\left[j_1+\frac{1}{2},j_2-\frac{1}{2},a+1,-j_1-j_2-1\right]}
\text{ for } j_2\geq \frac{1}{2}\nonumber
\eeqa
\end{itemize}
Explicit expressions for the atypical branching functions are now
obtained by plugging in our formula \eqref{branchingKac1} for the
branching functions of Kac modules. The coefficients of atypical
branching functions turn out to be positive.

\section{Vanishing invariants on $\CPn$}\label{sec:vanishing_invariants}

We start by considering a general symmetric superspace $G/H$, where $G$ is a Lie
supergroup with an involutive automorphism $\sigma$ such that $H$ is the maximal
compact subgroup of $G$ fixed by $\sigma$.
Let $e$ be the identity of $G$ and consider the point $o=eH$.
The Riemann structure on $G/H$ is defined by the requirement that $G$ is a
supergroup of isometries.
This means that the action of $G$ defines the metric and the curvature tensor
globally once their values are given at a single point, say $o$.

Let now $\mathfrak{g}$ and  $\mathfrak{h}$ denote the Lie superalgebras
of the Lie supergroups $G$ and $H$ respectively.
Define the quotient vector space $\mathfrak{m}=\mathfrak{g}/\mathfrak{h}$.
The commutation relations of $\mathfrak{g}$ split with respect to the involutive
automorphism $\sigma$ into the following three families
\begin{equation}
    [\mathfrak{h},\mathfrak{h}]\subset \mathfrak{h},\qquad
    [\mathfrak{h},\mathfrak{m}]\subset \mathfrak{m},\qquad
    [\mathfrak{m},\mathfrak{m}]\subset \mathfrak{h}.
\label{eq:m_rep}
\end{equation}
In particular, this means that $\mathfrak{m}$ is a representation of
$\mathfrak{h}$, which we denote by $\rho:\mathfrak{h}\rightarrow
\gl(\mathfrak{m})$.

The curvature tensor for symmetric spaces
\begin{equation}
    R_o(X,Y) Z =  [[X,Y],Z],\qquad X,Y,Z\in \mathfrak{m},
\label{eq:curv_tensor_sym_spaces}
\end{equation}
was computed in \cite{Helgason}.
We straightforwardly generalize this expression to symmetric superspaces, as
long as $X,Y,Z$
are even graded vectors.
Let $\beta$ be a $\mathfrak{g}$-invariant, non-degenerate, supersymmetric and
consistent
form on $\mathfrak{g}\times\mathfrak{g}$.
If $\mathfrak{m}$ is an \emph{irreducible real} representation of
$\mathfrak{h}$,
then the  solution to the condition that
$H$ is an isometry group
\begin{equation*}
    (h\cdot X,h\cdot Y)_o = (X,Y)_o,\qquad X,Y\in\mathfrak{m}
\end{equation*}
is uniquely determined, up to a proportionality constant called the radius of
$G/H$, by the restriction of $\beta$ to
$\mathfrak{m}\times\mathfrak{m}$
\begin{equation}
    (X,Y)_o = \beta(X, Y).
\label{eq:metric_bla}
\end{equation}
Note that, in order to be compatible with the automorphism
$\sigma$, the invariant $\mathfrak{g}$-form $\beta$
must be block diagonal with respect to the direct sum decomposition
$\mathfrak{g}=\mathfrak{h}\oplus\mathfrak{m}$.
Therefore,  the non-degeneracy of $\beta$ implies the non-degeneracy of
$(\phantom{x},\phantom{x})_o$ as defined in eq.~\eqref{eq:metric_bla}.

The curvature tensor being covariantly constant, it commutes with the action of
$H$ at $o$.
It will prove more comfortable to use instead of this commuting homomorphism
\begin{equation*}
    R_o\in \Hom_{\mathfrak{h}}\big(\wedge^2 \mathfrak{m}\otimes\mathfrak{m},
\mathfrak{m}\big)
\end{equation*}
the endomorphism $\Omega_o\in \End_{\mathfrak{h}}\mathfrak{m}\otimes
\mathfrak{m}$
defined the following way
\begin{equation*}
    \big(Y\otimes W, \Omega_o(Z,X)\big)_o = \big(W,R_o(X,Y)Z\big)_o =
\big([X,Y],[Z,W]\big)_o,
\end{equation*}
where the scalar product on $\mathfrak{m}\otimes\mathfrak{m}$ is defined as
\begin{equation*}
    (X\otimes Y, Z\otimes W)_o = (W,X)_o (Y,Z)_o.
\end{equation*}
Let $T_i$ be a basis of $\mathfrak{m}$ and $T_a$ be a basis of $\mathfrak{h}$.
Again, because $\beta$ is block diagonal
with respect to the decomposition $\mathfrak{g}=\mathfrak{m}\oplus\mathfrak{h}$,
the restriction of $\beta$ to $\mathfrak{h}\times \mathfrak{h}$ is
non-degenerate.
Denote by $T^a$ the basis dual to $T_a$ with respect to
$\beta$, that is
\begin{equation*}
    \beta(T^a,T_b) = \delta^a_b.
\end{equation*}
We shall rise and lower the group indexes with the help of the form $\beta$
and its inverse rather than
with the Killing form of $\mathfrak{g}$, which might be degenerate even for
simple Lie superalgebras.
Because of eq.~\eqref{eq:metric_bla}, this is consistent with the rising and the
lowering
of tensor indexes at $o$ with the metric $(\phantom{x},\phantom{x})_o$ and its
inverse.
Using the eqs.~\eqref{eq:m_rep} one can show that
\begin{equation*}
    \Omega_o (X,Y) = (-1)^{|a|}[T^a, X]\otimes [T_a, Y].
\end{equation*}
Put differently, the previous equation can be written as
\begin{equation*}
\Omega_o = (-1)^{|a|}\rho(T^a)\otimes \rho(T_a) = \rho(T_a)\otimes \rho(T_b)
\beta^{ab},
\end{equation*}
where
\begin{equation*}
    \beta_{ab}=\beta(T_a,T_b)
\end{equation*}
and $\beta^{ab}$ is the inverse of $\beta_{ab}$.
It becomes now obvious that a non-zero contraction in a tensor power of
$\Omega_o$
\begin{equation*}
    \Omega^{\otimes n}_o = \rho(T_{a_1})\otimes \rho(T_{a_2})\otimes \cdots
    \otimes \rho(T_{a_{2n-1}})\otimes \rho(T_{a_{2n}}) \beta^{a_2a_1}\cdots
\beta^{a_{2n}a_{2n-1}}
\end{equation*}
will result in a fusion of the type
\begin{equation*}
    \rho(T_{a_i})\otimes \rho(T_{a_j})\rightarrow \rho(T_{a_i}T_{a_j}).
\end{equation*}
In particular, subtracting all but one trace in $\Omega_o^{\otimes n}$ one gets
an expression
of the form
\begin{equation}
    \rho(T_{a_1}\cdots  T_{a_{2n}})\beta^{a_{2n}\cdots a_1},
\label{eq:all_but_one_trace}
\end{equation}
where $\beta^{a_{2n}\cdots a_1}$ is one of the $(2n-1)!!$
$\mathfrak{h}$-invariant tensors that can be constructed by raising to the
$n$-th tensor power
the $\mathfrak{h}$-invariant tensors $\beta^{a_ia_j}$.
Denote by $\mathcal{Z}(\mathfrak{h})$ the center of the universal enveloping
superalgebra
$\mathcal{U}(\mathfrak{h})$ of $\mathfrak{h}$.
Then we see that the expression in eq.~\eqref{eq:all_but_one_trace} is an
element of
$\mathcal{Z}(\mathfrak{h})$ in the representation $\rho$.
We arrive at the conclusion that all
$\mathfrak{h}$-invariant rank 2 tensors built from the tensor powers of the
curvature tensor $R_o$
by tracing the appropriate number of times with the metric
$(\phantom{x},\phantom{x})_o$ can be interpreted as elements of
$\mathcal{Z}(\mathfrak{h})$ in the
representation $\rho$.

Consider now the case of complex projective superspaces
\begin{equation*}
    \CPn = \UU(S|S)/\UU(S-1|S)\times \UU(1).
\end{equation*}
Complexifying everything, we get that $\mathfrak{m}$
is the direct sum of the fundamental representation
$\square_{S-1|S}$ of $\ssl(S-1|S)$ and of its conjugate $\bar{\square}_{S-1|S}$, thus revealing the complex structure of the supermanifold.
Moreover, $\mathfrak{h}=\ssl(S-1|S)\oplus\mathfrak{z}$, where $\mathfrak{z}$ is a two dimensional
center.
Let $\beta$ be the $\gl(S|S)$-invariant, non-degenerate
form provided by the supertrace in the fundamental representation.
Then the restriction of  $\beta$ to $\mathfrak{h}\times \mathfrak{h}$ is block diagonal with respect
to the direct sum decomposition $\mathfrak{h}=\ssl(S-1|S)\oplus\mathfrak{z}$.
One can choose as basis for $\mathfrak{z}$ the central element $E$ of $\gl(S|S)$ together with
its dual $N$ with respect to $\beta$.
Recalling that the invariant tensor $\beta^{a_{2n}\cdots a_1}$ were built from
tensor products of $\beta^{a_ia_j}$, we notice that $E$ and $N$ can only appear in eq.~\eqref{eq:all_but_one_trace}  in pairs.
Therefore, given that $E$ is in the kernel of $\rho$, the invariant tensors in eq.~\eqref{eq:all_but_one_trace} are effectively in the $\rho$-image of $\mathcal{Z}\big(\ssl(S-1|S)\big)$.
Finally, all these must vanish
because $\square_{S-1|S}$ and $\bar{\square}_{S-1|S}$ both belong to the block of the trivial representation of $\ssl(S-1|S)$.

\end{appendix}

\end{document}

%% file: table.dat
1 & 0 & 1.050128 & 1.037253 & 1.010766 \\ 
2 & 0 & 1.098296 & 1.094754 & 1.070405 \\ 
0 & -1 & 0.98817 & 0.969892 & 0.945022 \\ 
0 & -2 & 1.016252 & 1.006706 & 0.984296 \\ 
1 & -2 & 1.034566 & 1.033275 & 1.014131